\documentclass{aa}  
\usepackage{graphicx}
\usepackage{amsmath}	
\usepackage{amssymb}	
\usepackage{relsize}
\usepackage{xcolor}
\usepackage{txfonts}
\usepackage[colorlinks=true,linkcolor=blue,allcolors=blue]{hyperref}%

\usepackage{graphicx}
\usepackage{caption}
\usepackage{subcaption}

\usepackage[colorlinks=true,linkcolor=blue,allcolors=blue]{hyperref}%

\usepackage{ulem}

\defcitealias{2024spiniello}{INSPIRE V}
\defcitealias{2018Scodeggio}{VIPERS}

\begin{document} 

\title{\texttt{ILLUSTR}ating red nugget assembly through observations and simulations}

   \author{Micheli T. Moura\inst{1,2}
   \thanks{E-mail: micheli.t.moura@gmail.com}
   \and
   Anna Ferré-Mateu \inst{1,3,4}
   \and
   Ana L. Chies-Santos 
   \inst{2}
   \and
   Cristina Furlanetto
   \inst{2}
   \and Michael A. Beasley \inst{1,3,4}
   }
   \institute{
Instituto de Astrofísica de Canarias, Vía Láctea s/n, E-38205 La Laguna, Tenerife, Spain\\
Instituto de Física, Universidade Federal do Rio Grande do Sul, Av. Bento Gonçalves 9500, 
Porto Alegre, R.S. 90040-060, Brazil\\
Departamento de Astrofísica, Universidad de La Laguna, E-38200 La Laguna, Spain\\
Centre for Astrophysics and Supercomputing, Swinburne University, John Street, Hawthorn VIC 3122, Australia\\
}

   \date{Received; accepted}
   
   \titlerunning{\textit{Illustrating massive relic galaxies}}
\authorrunning{Micheli, T. Moura. et al.}  
 
  \abstract
   {
The properties of massive and compact early-type galaxies provide important constraints on early galaxy formation processes. Among these, massive relic galaxies, characterized by old stellar populations and minimal late-time accretion, are considered to be preserved compact galaxies from the high-$z$ Universe. In this work, we investigate the properties of compact and massive galaxies (CMGs) using the TNG50 cosmological simulation, applying a uniform selection criteria that matches observational surveys at $z=0$, $z=0.3$, and $z=0.7$. This allows for direct comparisons with observed compact galaxies at each evaluated redshift. We classify CMGs according to their stellar mass assembly histories to investigate how compactness relates to dynamical properties and chemical enrichment across cosmic time. Our results show that simulated CMGs consistently follow the observed mass--size relation, with the number of compact galaxies increasing at higher redshifts; this number density follows the trend in observational data. In the dynamical context, while observations suggest that relic galaxies are outliers in the stellar mass--velocity dispersion plane, the simulated compacts show relatively uniform velocity dispersions across different accretion histories. Observed relics tend to be more metal-rich than other compact galaxies with extended star formation histories, deviating from the local mass--metallicity relation. In contrast, simulated compact galaxies are, overall, all more metal-rich than the quiescent population, regardless of their accretion histories. We also find that the deviation of the simulated CMGs from the mass-metallicity relation decreases with increasing redshift. These findings suggest that the extreme characteristics of CMGs in TNG50, particularly regarding metal enrichment and dynamical properties, are less pronounced than those observed in real relic galaxies. Nonetheless, the results offer a theoretical framework to assess the properties of such extreme objects from different epochs, highlighting both alignment with and deviations between the models.

}

\keywords{galaxies: structure -- galaxies: dynamics -- galaxies:observations  -- galaxies: relic galaxy}

   \maketitle
%
\section{Introduction}
\label{sec:intro}
The two-phase scenario (e.g.,\,\citealt{2009Naab,2010OserOstriker}) for the formation and evolution of massive Early-Type Galaxies (ETGs) has gained a new perspective with the advent of the JWST data. The scenario is described by an initial phase of formation occurring before $z=2$, which is marked by the rapid growth of stellar mass through wet mergers, resulting in a phase with a high star formation rate (SFR), but without significant size growth (e.g.,\,\citealt{2015ApJVanDokkun,2015Zolotov,2020Zibetti}). Due to the rapid mass increase and the intense star formation episodes in this first stage of evolution, a massive and compact object is formed, known as red nugget. They are expected to quickly quench and become passive. However, the recent discoveries from JWST have revealed that quenching in massive galaxies was far more common during the first two billion years ($z > 3$), leading to higher-than-expected number densities of massive quiescent galaxies in the redshift range $3 < z < 5$ \citep{2024Carnall,2024valentino,2024Long}. These high-$z$ red nuggets will subsequently undergo a second phase, whereby they grow in size through the accretion of smaller satellites, creating the massive ETGs we see nowadays in the local Universe. 

Following this scenario, compact massive galaxies (CMGs) are thus expected to decrease in number density with decreasing redshift (e.g.,\,\citealt{2009Trujillo, 2015Damjanov, 2017Charbonnier}), with the underlying expectation that CMGs in the local Universe should be very rare, as the majority will grow in size by mergers and accretion during the second phase. Nonetheless, local CMGs have been found in all environments (e.g.,~\citealt{2009Trujillo, 2013Poggianti, 2023GrebolTomas}). Their existence is explained by the fact that mergers are stochastic, and thus it is expected that a small fraction of the high-$z$ red nuggets will avoid significant accretion \citep{2013Quilis} over the second phase. In those extremely rare cases, where the red nugget has remained untouched over cosmic time, such CMGs have been named \textit{massive relic galaxies} \citep{2009Trujillo}.

The search for these elusive types of ancient galaxies has revolved around small samples of local CMGs. Unfortunately, the great majority of them turned out to be younger systems \citep{2009Trujillo,2012Ferre-mateu,2013Poggianti}. Some CMGs were promising, showing several characteristics similar to massive relics (e.g., \citealt{Yildrim2017,2023GrebolTomas,2024Clerici}). However, only a few have been thoroughly investigated through detailed multi-property analyses. Among them, PGC\,032873, Mrk\,1216, and NGC\,1277 stand out as the most extensively studied cases of massive relic galaxies \citep{Ferre-Mateu2017}. Their disk-like symmetrical shapes show no signs of tidal streams, indicating no recent evident interaction signatures; all host extremely large supermassive black holes \citep{2015Ferre-mateu}, present a bottom-heavy IMF \citep{2015Martinnavarro,Ferre-Mateu2017}, and have stellar populations and kinematics that can be compared with high-$z$ red nuggets. The luminosity fraction of the hot inner stellar halo (or spheroid compactness) correlates strongly with merger history \citep{2025Zhu,2022ZhuLing}, with merger-free galaxies showing extremely compact spheroids, consistent with those seen in relic candidates \citep{2024Moura,2025Zhu}. In recent years, however, new promising relic candidates have begun to emerge, supported by improved photometric selection techniques, deeper spectroscopic follow-up, and dedicated kinematics analysis (e.g., \citealt{2023GrebolTomas,2025Zhu,2025Mills}).

While massive relic galaxies are understood within the proposed theoretical scenario, it is yet unclear how the younger CMGs can exist in the local Universe. To investigate this, \citet{2023GrebolTomas} searched for CMGs in MaNGA (DR17,\,\citealt{2015ABundy}) and classified them using machine learning approaches, according to their star formation histories (SFH), i.e., how fast and when they occurred. They found that CMGs can be classified into three groups, one of which highly resembles massive relics. The other types show distinctly different SFHs (and also metallicities and $\alpha$-abundances). Some show extended SFHs similar to lower mass ETGs, while some present fast formation, although delayed in time (late-bloomers). More recently, \citet{2025Mills} have searched for CMGs in the SDSS, showing that indeed these objects exist by larger numbers than seen before. With the upcoming all-sky surveys such as {\textit Euclid}, we expect that the number of CMGs and known relics will change.

As expected from the two-phase scenario, a significant increase in the number density of CMGs between $z = 0$ and $z \sim 2$ occurs as the galaxy has had less time for the accretion phase \citep{2009Damjanov,2017Charbonnier,2018Tortora,2023Lisiecki}. This means that massive relic galaxies should also exist at intermediate redshifts. The \textsc{INSPIRE} project \citep{2021INSPIREproject} has focused on investigating CMGs at intermediate redshifts to create the first catalogue of spectroscopically confirmed massive relic galaxies between $0.1 < z < 0.5$. This project has so far provided a detailed characterization of the kinematics, stellar populations, the low-mass end of the initial mass function (IMF), and environment \citep{2023ADago,2024spiniello,2023MartinNavarro,2024Maciata,2024Scognamiglio} of CMGs in that redshift range. One key finding from this project is that massive relic galaxies tend to be more metal-rich compared to non-relics, while also exhibiting higher velocity dispersions \citep{2023ADago}, trends also seen in the local CMGs of \citet{2023GrebolTomas}.

Following the idea that a `degree of relicness' (DoR) seems to exist in local massive relics \citep{Ferre-Mateu2017} and given the three clearly distinctive types of CMGs from \citet{2023GrebolTomas}, the CMGs from the \textsc{INSPIRE} project were classified according to a parametrization of the DoR \citep{2024spiniello}. It was defined as a dimensionless value ranging from 1 to 0, where 1 represents the most extreme case of a relic galaxy (e.g., NGC\,1277 in the local Universe is the prime example of an extreme relic, with a DoR of almost 1), while 0 represents a galaxy still forming stars. They found that the DoR correlates with stellar velocity dispersion, metallicity, [Mg/Fe] ratio, and age. In addition, \citet{2024Scognamiglio} observed that CMGs could reside both in clusters and in the field similar to what has been found in the local Universe (e.g.,\,\citealt{2013Poggianti,2016PeraltaArriba,2014mariacebrian,2024Moura}). However, there is a slight preference for galaxies with a very low DoR ($< 0.3$) to be found in under-dense environments, while the most extreme relics tend to be found in the central regions of clusters. 

Going to even higher redshifts, the VIMOS Public Extragalactic Redshift Survey (VIPERS, \citealt{2018Scodeggio}), identified 77 CMGs in the redshift range $0.5 < z < 1.0$. This catalogue provides spectroscopically confirmed galaxies that follow the mass--size and quiescence criteria for red nuggets at this intermediate redshift \citep{2023Siudek,2023Lisiecki}. \citet{2023Lisiecki} emphasizes the impact of source compactness limits on selection, with sample size varying by up to two orders of magnitude depending on the criterion. It is expected also that the fraction of red nuggets in clusters may increase over time as red massive normal-sized galaxies lose their envelopes through stripping \citep{2023Siudek,2025Zhu}. The resulting number density found in VIPERS exceeds estimates for the local Universe but remains below values at $z>2$, bridging the gap at intermediate redshifts.

The properties of CMGs, in particular massive relic ones, have also been discussed in several numerical simulations \citep{2015Wellons,2015Stringer,2016PeraltaArriba,Floresfreitas2022,2024Moura,2025Zhu}. Using the Illustris TNG50 simulation (e.g.,\,\citealt{2019Pillepich,2019Nelson}), \citet{Floresfreitas2022} identified five massive relic galaxy candidates that match observed ones in terms of metallicity and morphology. Also employing TNG50 simulations, \citet{2024Moura} analysed a sample of CMGs, identifying relics as those with less than 10\% of satellite accretion throughout their evolution. Their analysis of internal dynamics and environment revealed that merger signatures can be detected through the kinematics of the inner stellar halo, providing a new signature for identifying extreme relics. \citet{2025Zhu} showed that in TNG50, the hot inner stellar halo fraction tightly correlates with merger history. A very low halo fraction combined with a dynamically cold disk signals a merger-free relic. They identified seven compact galaxies from \citet{Yildrim2017} consistent with this scenario.

Fortunately, cosmological simulations have the advantage of being able to investigate the red nugget progenitors of the local CMGs and tracing back their evolutionary pathways. In this work, we us the TNG50 simulation to select CMGs at four redshift bins using the similar primary selection criteria applied to observations. This approach allows for a consistent comparison of stellar populations and kinematics between simulated and observed CMGs at different epochs, including $z=0$ for local relations, $z=0.3$ (for comparison with INSPIRE), and $z=0.7$ (for comparison with VIPERS). Although no red nugget catalog is available for $z\sim2$, we include this redshift bin because it has been traditionally adopted as the reference point for the onset of the second-phase evolutionary scenario. Recent JWST results, however, suggest that compact massive galaxies could already be present at earlier epochs ($z\sim3$–4; \citealt{2023Carnall,2025Nana}), indicating not necessarily an earlier onset of their formation, but rather that some systems may have formed much more rapidly than previously thought. For consistency with the established two-phase framework, we therefore use $z=2$ as our starting point.

Our goal is to draw a parallel between observed and simulated CMGs by examining their mass--size, mass--velocity dispersion, and mass--metallicity relations, highlighting the aspects as compactness, dynamics, and chemical abundances across different cosmic epochs and their relation with different assembly histories. This paper is organized to introduce the simulation sample selection and observational data in Section \ref{sec:sampleselect}. Analysis and discussion are presented in Section \ref{sec:results}, and a summary and conclusions are shown in Section \ref{sec:conclusions}. Throughout the paper, we assume the \citet{plank2016} parameters to $\Lambda$CDM cosmological model with $H_{0} = 67.74\,\rm{km}\,\rm{s}^{-1} \rm{Mpc}^{-1}, \Omega_{m} = 0.30$, and $\Omega_{\Lambda} = 0.69$.

\begin{figure*}[h!]
\centering
\includegraphics[width=\textwidth]{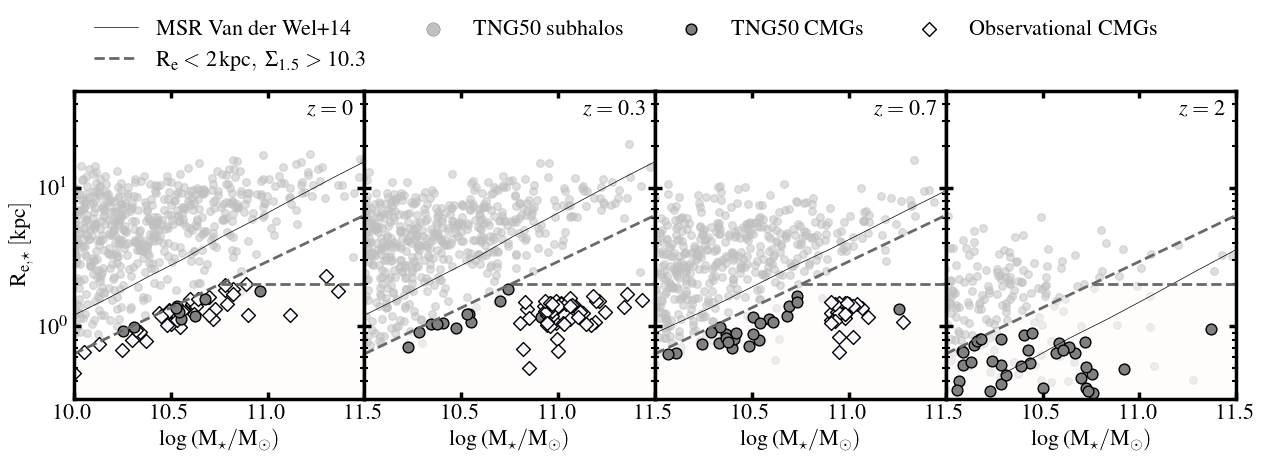}
\caption{Stellar mass--size relation across redshift bins. All TNG50 subhalos are shown as light gray circles in the background, while selected compact subhalos are represented by dark gray circles. Observational CMGs from the different samples are indicated with white diamond markers. The solid black lines follow the mass--size relation from \citet{vanderWel2014}, while the dashed gray lines highlight the selection region for the simulated sample ($R_e<2\,\rm{kpc}, log\,\Sigma_{1.5}>10.3\,dex$)}.
\label{fig:mas-size}
\end{figure*}
%
\section{Sample selection and main properties}
\label{sec:sampleselect}

We explore the formation and evolution of CMGs by analyzing the SFHs from simulations and observations at four redshifts: $z=0$, $z=0.3$, $z=0.7$, and $z=2$. We aim to assess how structural properties, such as stellar mass, size, metallicity, and velocity dispersion, relate to the accretion histories, providing insight into their evolutionary pathways. Here we describe the parameters employed to select the sample and the main properties analyzed.

\subsection{Observations}

Traditionally, CMGs have been sought by applying a mass and size criteria. Although not all works have followed the exact same criteria (see \citealt{2017Charbonnier} and \citealt{2023Lisiecki} for discussion), the initial, most restrictive criteria from \citet{2009Trujillo} defined CMGs by having $M_\star \ge 8\times10^{10}\,\rm{M}_\odot$ and $R_e \le1.5\,\rm{kpc}$. Later works, such as INSPIRE, have allowed for a slightly larger size constraint limiting them at 2\,kpc. Some other later works also incorporated a third criteria, based on the density surface relation established by \citet{2013Barro}. This compactness parameter is defined as $\Sigma_{1.5} = \rm{M\,[M_\odot]/(\rm{R_e\,[kpc])^{1.5}}}$, where one considers the stellar mass contained within the effective radius. Here we use $\Sigma_{1.5}>10.3$\,dex for the selection \citep{2013Barro, 2023GrebolTomas, 2025Mills}. This relation provides a compactness parameter consistent across all redshifts adopted.

For the local Universe sample, we consider on one side the three confirmed massive relic galaxies: NGC\,1277, PGC\,032873, and Mrk\,1216 \citep{Ferre-Mateu2017,2014Trujillo}. All the structural properties, such as stellar masses, stellar population parameters, and kinematics, have been taken from \citet{Ferre-Mateu2017}. We also include the 37 MaNGA CMGs from \citet{2023GrebolTomas}, obtained by selecting galaxies with surface mass density threshold of $\Sigma_{1.5}>10.3$\,dex and $\rm{R_e<2}\,\rm{kpc}$. Using stacked spectra within $1\,\rm{R_e}$, they measured recessional velocities, stellar velocity dispersions, and derived stellar population properties, including age and total metallicity. Through the analysis of the stellar population and kinematic properties, they classified the galaxies into three groups based on their assembly history: Group A with very old galaxies ($>12$\,Gyr) with early and steep SFHs (similar to massive relics); Group B with intermediate-ages and more extended SFHs ($\approx 8$\,Gyr); and Group C, young galaxies ($\approx 5$\,Gyr), named `late-bloomers' as they formed fast but late in cosmic time.

On local-to-intermediate redshift, we use the entire sample from the INSPIRE project \citep{2021INSPIREproject}, consisting of 52 CMGs at redshifts between 0.1 and 0.5, all with effective radii smaller than 2\,kpc and stellar masses greater than $6 \times 10^{10}~\rm{M_{\odot}}$. This sample includes optical and near-infrared photometry from the KiDS and VIKING surveys \citep{2013Edge}, structural parameters \citep{2018Roy}, and stellar masses inferred through spectral energy distribution (SED) fitting using the \textit{u, g, r,} and \textit{i} bands \citep{2018Tortora,2020AScognamiglio}. The INSPIRE galaxies have estimates of integrated stellar velocity dispersions within $1\,\rm{R_e}$, [Mg/Fe] measurements obtained from line indices, as well as mass-weighted stellar ages and metallicities derived from full spectral fitting. This sample, similarly to the MaNGA one, shows a variety of SFHs, parametrized by a varying DoR \citep{2024spiniello}.

Finally, we use data from the VIMOS Public Extragalactic Redshift Survey (VIPERS, \citealt{2018vimos}) for the intermediate redshift regime. For this work we use a subset of 77 CMGs from the VIPERS catalogue at  $0.5 < z < 1.0$ \citep{2023Lisiecki}, constructed by selecting objects both massive and compact ($\rm{M}_\star > 8 \times 10^{10} M_{\sun}$, $\rm{R}_{e}< 1.5$ kpc) but also including passiveness criteria, based on both spectroscopic and photometric analyses (see \citealt{2023Lisiecki} and \citealt{2023Siudek}). 

All observational data described here are shown in Fig.~\ref{fig:mas-size}, represented by diamond-shaped markers, separated by redshift bins. The figure also presents the stellar mass--size relation of observed galaxies from \citet{vanderWel2014} at each redshift of interest.

\subsection{Simulations}
\label{subsec:simulations}
We use the IllustrisTNG suite of cosmological magnetohydrodynamical simulations \citep{2018Marinacci,2018Naiman,2018Nelson,2018Pillepich,2018Springel} to select our sample of simulated CMGs. These simulations were performed using the moving-mesh code \textsc{Arepo} \citep{2010Springel}, with a comprehensive overview provided in the release paper by \citet{2019Nelson}. In this work we will focus on TNG50, (e.g.,\,\citealt{2019Pillepich,2019Nelson}), the smallest cosmological volume in the suite ($\rm{L}_{\rm box} = 51.7~\rm{cMpc}$), which offers the highest resolution \citep{2018Pillepich}. While a smaller volume will result in a small number of candidates, the higher resolution allows us to study the structural and dynamical properties to compare with observations.

Stellar masses are computed as the total mass of bound star particles within a given aperture, while galaxy sizes are typically defined based on the three-dimensional half-mass radius of the stellar distribution. The radius, in this context, corresponds to the half-mass radius, which encloses half of the total stellar mass of the galaxy. Adopting the 3D half-mass radius provides a consistent and physically motivated definition of size within the simulation, ensuring that the selected sample is genuinely compact, regardless of the observational size definition. For further discussion on mass–size comparisons between simulations and observations, see \citet{2018Genel}.
To select our simulated sample, we selected all subhalos with stellar masses above $10^{10}\,\rm{M_\odot}$ and sizes $\rm{R_e}\,<2\,\rm{kpc}$ to compare with the observational sample described before. We use \citet{2013Barro} surface density to select all the compact subhalos among this first selection. The final selection of subhalos can be seen in the mass--size plane in Fig.~\ref{fig:mas-size}, compared to the observational data. It is important to note that this selection was applied independently at each redshift ($z=0$, $0.3$, $0.7$, and $2$), without following the merger trees of the subhalos. Thus, while some galaxies selected as compact at higher redshift may correspond to progenitors of $z=0$ galaxies, this is not systematically tracked or required by our selection. Our focus is to obtain the population of CMGs at each cosmic epoch based solely on their properties at that specific snapshot, independent of whether their descendants or progenitors remain compact.

We obtain the same mass range for MaNGA galaxies and the subhalos at $z=0$, with the most massive galaxies at this redshift being the three spectroscopically confirmed relics. At $z=0.3$ and $z=0.7$, the stellar mass range for observed compacts is in general concentrated in the high-mass end ($\rm{M_\star}\approx10^{11}\,\rm{M_\odot}$), while the simulated compacts have on average lower stellar masses, when compared with the observed ones. The simulated galaxies may not reach this high-mass end due to the limited volume of the simulation. We observe an increase in the number of subhalos and compact galaxies as redshift increases. This is due to the mass--size evolution, where, at early times, galaxies tend to be more compact than they are nowadays. 

\begin{figure}
\centering
\includegraphics[scale=0.35]{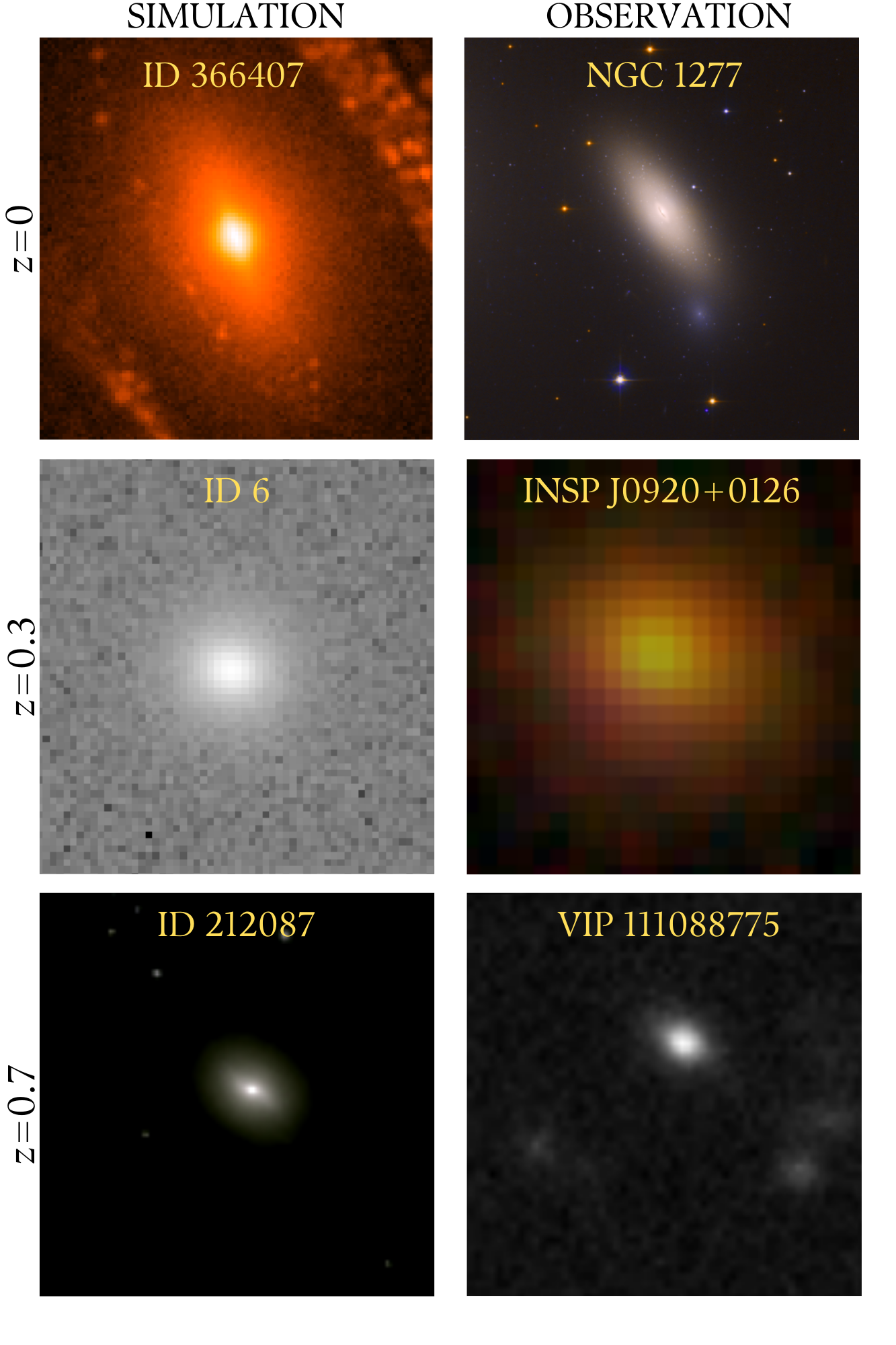}
\caption{Example of the selected sample of compact TNG50 simulated (left column) and observed (right column) galaxies across redshift bins: $z=0$ (top row), $z=0.3$ (middle row), and $z=0.7$ (bottom row). Observed galaxies include NGC\,1277 (top-right, HST ACS/WFC instrument), \citealt{2024spiniello} (middle-right), and \citetalias{2018Scodeggio} (bottom-right).}
\label{fig:mont}
\end{figure}

In addition to the mass--size and compactness selection criteria, we also analyzed the morphology of the simulated galaxies through visual inspection to ensure that none of them had clear merger signatures. After this, we obtained a final sample of nine compact massive subhalos at $z=0$, 13 subhalos at $z=0.3$, 25 compact subhalos at $z=0.7$, 36 subhalos that match the mass--size relation and concentration parameter at $z=2$ (to represent the red nugget stage, i.e. the start of the second-phase). An example of a subhalo at each redshift bin from the simulated sample is shown in Fig.~\ref{fig:mont}, alongside real compact galaxies at the same redshift.  

\begin{figure}
\centering
\includegraphics[width=\columnwidth]{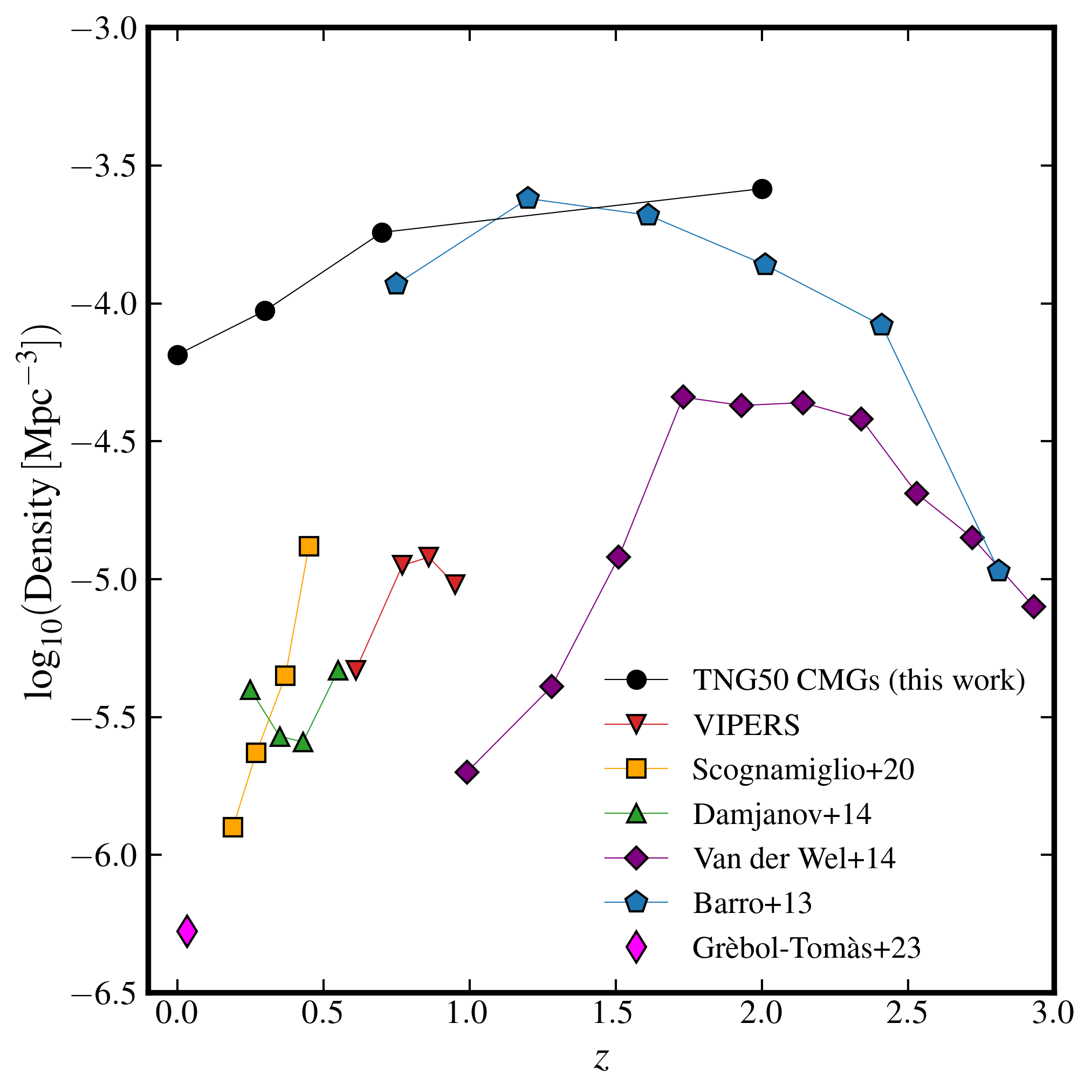}
\caption{Number density of CMGs as a function of redshift, in units of co-moving Mpc$^{-3}$. The black line and markers represent this work TNG50 CMGs. Colored lines and markers correspond to observational estimates from VIPERS \citep{2023Lisiecki} in red, \citet{2020AScognamiglio} in orange, \citet{2014Damjanov} in green, \citet{vanderWel2014} in purple, \citet{2013Barro} in blue, and \citet{Ferre-Mateu2017} in dark green for the local relics, and in magenta for \citet{2023GrebolTomas} compacts. Simulated CMGs show increasing number density with redshift, in line with models predicting more compact objects in the early Universe. A similar trend is seen in the observational data.}

\label{fig:ndensity}
\end{figure}

Our simulated sample shows an increased number of selected CMGs with increasing redshift. By calculating their number density across redshift bins at Fig.~\ref{fig:ndensity}, we demonstrate agreement with the established trend of increasing compact galaxy abundance at comparable redshifts, as reported by \citet{vanderWel2014} and \citet{2013Barro}. We limited our selection to $z \leq 2$ since our analysis focuses on this redshift interval. This also enables a direct comparison with observational estimates from other works on compact galaxies (e.g.,\,\citealt{2023Lisiecki,2020AScognamiglio,vanderWel2014,2013Barro,Ferre-Mateu2017}). It is important to emphasize that the compact selection criteria varies across different studies (for a discussion, see \citet{2017Charbonnier,2023Lisiecki}). In addition, since number density depends on the survey volume, the limited size of the TNG50 simulation also affects the results. Therefore, direct expectations regarding the agreement between simulated and observed number densities at each redshift should be interpreted with caution.

The stellar velocity dispersion values ($\sigma_\star$) of simulated galaxies in TNG simulations are calculated based on the particle distribution within each subhalo. The \textsc{subfind} algorithm \citep{2010Springel,2009Dolag} uses the spatial $x$, $y$, $z$ center positions, and central velocities to compute the mass-weighted 3D velocity dispersion for each particle type (i.e., dark matter, gas, and stellar particles). Regarding these measurements, \citet{2024Sohn} investigated the stellar velocity dispersion of quiescent massive subhalos in IllustrisTNG, focusing on potential systematics. They analyzed the dependence of velocity dispersion on simulation resolution, viewing axes, and the identification of member stellar particles. The findings revealed no significant discrepancies in $\sigma_\star$ across these variations. Additionally, the study showed that the scatter in the velocity dispersion distribution is generally larger than the systematic uncertainties. Interestingly, they also found that in the $\rm{M}_\star-\sigma_\star$ relation, observed values from SDSS DR12 \citep{2015Alam} data at $z<0.2$ are generally 40\%–50\% higher than the simulated $\sigma_\star$ values for a given stellar mass. We show in the Appendix Fig.~\ref{fig:hist} a histogram with the distribution coverage in stellar mass, effective radius, and stellar velocity dispersion of both simulated and observed samples, and a Table \ref{tab:ids} with all the information on the simulated sample is also shown in Appendix.

Regarding the total metallicities, TNG (and Illustris) simulations model astrophysical processes such as radiative cooling and heating of the interstellar medium (ISM), stellar feedback, black hole growth, and active galactic nuclei (AGN) feedback, which include models of stellar evolution and chemical enrichment. In such models, star formation and pressurization of the multiphase ISM are treated following \citet{2003Sringel}, where gas exceeding a density threshold ($ n > 0.13\,\rm{cm}^3$) forms stars based on a \citet{2003Chabrier} initial mass function (IMF), with the dense, star-forming ISM treated using a two-phase effective equation of state model \citep{2018Pillepich}. The stars that are formed adopt their metallicity from the gas from which they form. As the stars evolve, they return both mass and metals to the ISM via supernovae and asymptotic giant branch stars according to tabulated mass and metal yields. The model that employs cell-centered magnetic fields to solve the equations of ideal magneto-hydrodynamics \citep{2011Pakmor} tracks the evolution of nine chemical elements (H, He, C, N, O, Ne, Mg, Si, Fe), for details on the implementation and metal tagging, see \citet{2018Pillepich} and \citet{2018Naiman}. The metallicity of each stellar particle in TNG is calculated as the ratio of the mass of heavy elements to the total mass of that particle at the time of its formation, reflecting the chemical composition of the gas cell it was born from. This intrinsic metallicity is stored as part of the stellar particle's properties and represents the local enrichment at the formation site.

\citet{2024Garcia} investigated the intrinsic differences between how metallicity is obtained in simulations and observations. They also analyzed the gas-phase and stellar metallicities in the Illustris, IllustrisTNG, and EAGLE \citep{2015Schaye,2015Crain} cosmological simulations to study their dependence on star formation and stellar mass. While these simulations share common approaches to modelling the ISM, star formation, and metal enrichment, they differ in how they detail the feedback processes that shape both the stellar mass-gas-phase metallicity relation and stellar-mass metallicity relation ($\rm{MZ_{\star}R}$). \citet{2024Garcia} used observational data from \citet{2005Gallazzi} and \citet{2008Panter} to contrast with the simulations, emphasizing that such comparisons are not straightforward.
Specifically, calculating metallicities by averaging the metal content of star particles weighted by their mass, as done in all three analysed simulations, results in higher stellar metallicities compared to the ones obtained using radiative transfer codes, which better approximate mock galaxy observation \citep{2016Guidi}. Their findings indicate that the $\rm{MZ_{\star}R}$ at $z=0$ and higher redshifts are more elevated compared to observational data. \citet{2018Nelsonn} noted that a robust comparison of TNG stellar metallicities with SDSS fiber-derived spectral values requires more advanced forward modeling. The study also pointed out that the stellar metallicities in the simulations do not align with observations, partly due to a shortage of present-day, low-mass dwarf systems formed within the past billion years, which could contribute to the observed discrepancy. However, more importantly, it suggests that this discrepancy is likely the result of an overly simplistic comparison. An alternative analysis using consistent fits to mock SDSS fibre spectra is necessary to enable a fair comparison between observations and simulations.

We note that we do not aim to directly quantitatively compare observations and simulations of CMGs because the methods used to derive parameters differ. Additionally, there is no equivalence regarding environment, universe volume, sample size, or resolution between observational and simulation data.

The most appropriate way to interpret this analysis — and any comparative analysis between observation and simulation — is to assess qualitatively whether the trends converge or diverge in general, considering the limitations and methodological differences in studying the same property. By grouping the samples according to their star formation histories, we evaluate the behaviour of the simulated populations in each redshift and whether they follow similar observed trends. In other words, we are not searching for simulated analogues of observed CMGs or comparing one-to-one values, but rather characterizing the populations in each dataset and comparing their general behavior. This perspective will guide our discussion of the results.

\section{Analysis}
\label{sec:results}
\subsection{Time scale of evolution}
\label{sec:timescale}

The two-phase scenario allows for different evolutionary pathways, where mergers can prolong the SFH by different degrees. Galaxies that assembled most of their mass early in cosmic history tend to exhibit more extreme properties, such as higher velocity dispersions, larger black hole masses, and over-solar metallicities, overall being seen as more extreme than galaxies that had an extended SFH. Given the diversity of formation timescales that massive galaxies show, a variety in the degree of `how much relic' a CMG can be also exists. Known as \textit{degree of relicness} (DoR; \citealt{Ferre-Mateu2017}), it has been explored and parametrized both for MaNGA and INSPIRE samples.

\citet{2023GrebolTomas} used machine learning tools to classify the SFHs of the MaNGA sample of CMGs into different types, accounting for both their SFHs and structural parameters. Applying a $K-$means algorithm, they found three distinctive groups: `Group A' was composed of very old, most compact, metal-rich galaxies, all showing rapid SFHs with a very early assembly, compatible with their high $\alpha$-enhancement values. The other two groups (B and C) showed different levels of extended SFHs, one being more similar to low-mass ETGs (i.e., showing steady extended SFHs), while the other presented a rapid SFH but extremely delayed in time (similar to the younger CMGs found in \citet{2012Ferre-mateu}, `late-bloomers'). Galaxies in groups B and C had overall lower metallicities, less enhanced $\alpha$-abundances, and lower velocity dispersions. 
\begin{figure}
\centering
\includegraphics[width=\columnwidth]{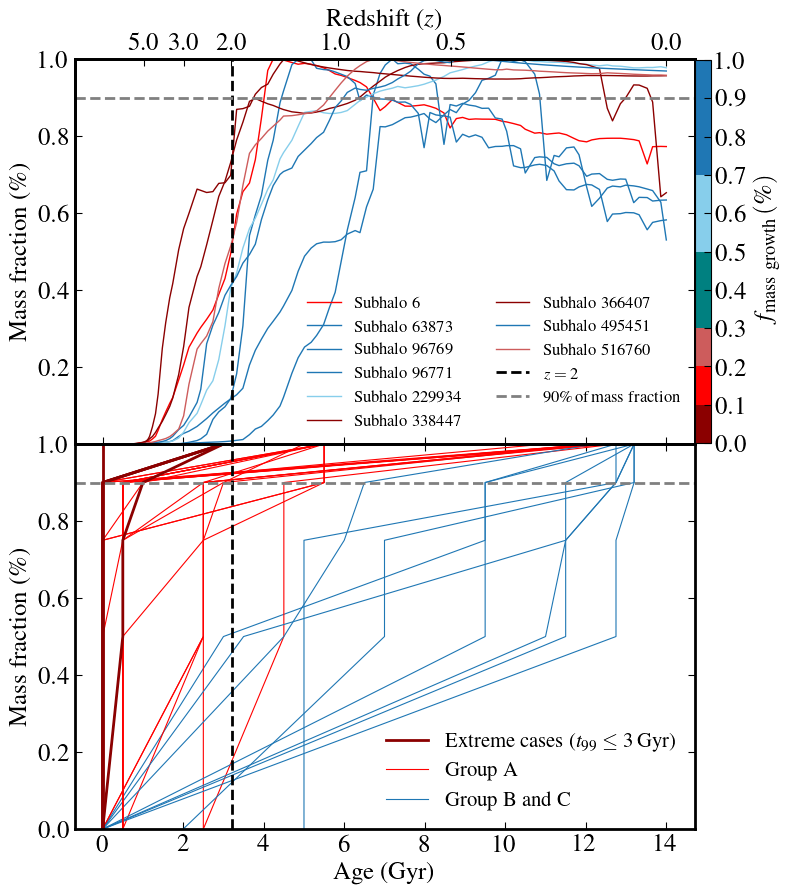}
\caption{Stellar mass assembly for the simulated and observed samples at $z=0$. The upper panel shows the stellar mass fraction over time for each compact subhalo selected from TNG50, color-coded by the stellar mass growth fraction since $z=2$. The lower panel shows the stellar mass assembly over time for the MaNGA compact galaxies of \citet{2023GrebolTomas}, color-coded by their groups: blue represents groups B and C, which show extended and/or late SFHs. Red color represents galaxies in their group A, which show early and fast formation (and thus negligible accretion after $z=2$), being dark red those extreme cases whereby the galaxy is fully assembled early on (and thus has the minimum amount of accreted material, $<10$\%). In both panels, the black dashed vertical line indicates $z=2$ and the dashed gray horizontal line indicates the stellar mass fraction of 90\%.}
\label{fig:assembly}
\end{figure}
In INSPIRE, galaxies from DR1 \citep{2021Spiniello} were classified into three categories based purely on their SFHs: \textit{extreme relics}, objects that fully assembled their stellar mass by $z=2$; \textit{relics}, objects that assembled at least 75\% of their stellar mass by $z=2$; and \textit{non-relics}, galaxies that had less than 75\% of their mass assembled by  $z=2$. This quantification of the DoR was later refined in the final DR5, by accounting for both the average SFH and the cosmic time when the majority of the stellar mass was assembled (see \citealt{2024spiniello}). A higher DoR (i.e., $>0.7$) indicates an earlier formation epoch, with minimal contribution from accreted stars or formed during later star-forming episodes. A lower DoR indicated that, although a portion of the stars are old and formed in early times, there is a significant presence of later-formed stellar populations with varying ages and metallicities, reflecting extended SFHs. For the \textit{eximium exemplum} case of NGC\,1277, its DoR is estimated to be 0.95 \citep{2024spiniello}. Nonetheless, the difference of using one parametrization over the other did not have an impact on the correlations derived with other stellar parameters such as metallicity, IMF, and stellar velocity dispersion.
 
Taking into account the different definitions of DoR, here we adopt a simplified version, with three distinctive categories. The \textit{relics} class, which are galaxies that show early and fast formation timescales, hereafter represented in red colors. This would correspond to galaxies on Group A of the MaNGA sample, and those with DoR $\ge 0.5$ in INSPIRE. Among this relic class, we select the \textit{extreme} cases as those that present the earliest and fastest formation timescales. This \textit{extreme} class is represented by the dark red color among the following figures and corresponds to the extreme cases of group A for MaNGA (see Fig.~\ref{fig:assembly}) and galaxies with DoR $\ge 0.7$ for INSPIRE data. Lastly, the \textit{non-relics} class comprises CMGs with a low likelihood of being relics given their late and/or more extended SFHs (DoR$<0.5$ and Groups B and C in MaNGA), which will be depicted in blue throughout the rest of the figures.

\begin{figure*}
\centering\includegraphics[width=\textwidth]{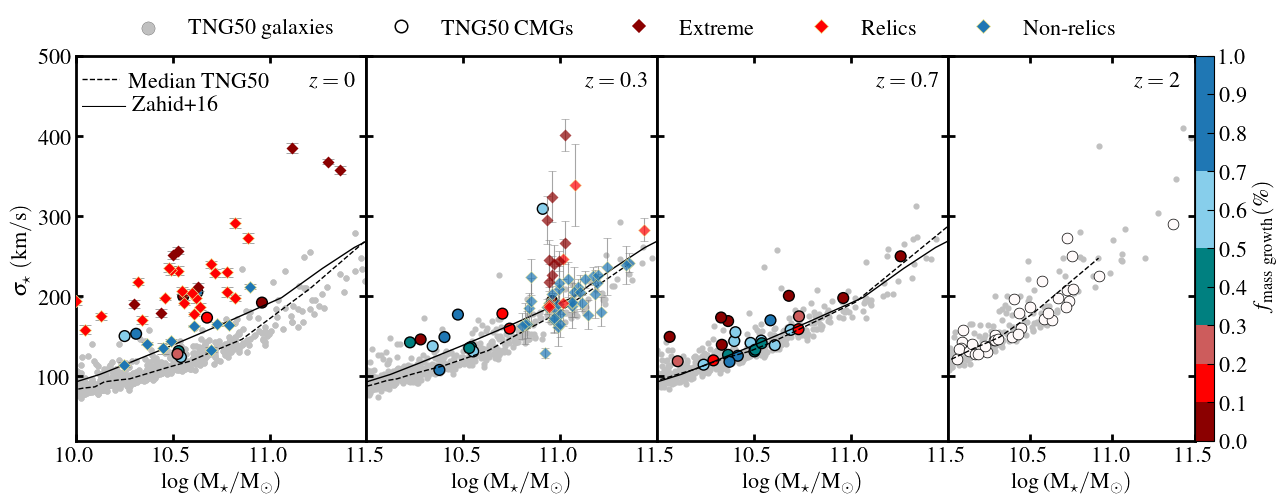}
\caption{Stellar mass--stellar velocity dispersion relations for simulated and observed galaxies in the different redshift bins. The dashed black line represents the median $\sigma_\star$ for all TNG50 quiescent subhalos, while the solid black line corresponds to the \citet{2016AZahid} relation across all redshifts. All TNG50 subhalos are shown as gray dots, being the compact sample color-coded by their mass growth since $z=2$ (except for the last bin, which is considered as the starting point and thus are left un-colored). Observational data are represented by diamonds color-coded as: dark red for the \textit{extreme} relics (e.g., extreme cases of group A for MaNGA, DoR$>$0.7 for INSPIRE); red for \textit{relics} (rest of A of MaNGA and INSPIRE galaxies with 0.5$<$DoR$<$0.7); and blue for the \textit{non-relic} group (classes B and C of MaNGA, DoR$<$0.5 for INSPIRE).}
\label{fig:sigma}
\end{figure*}

For the simulations, we use a similar approach, as it is possible to reconstruct the evolution of stellar mass over time by employing the \textsc{sublink} merger tree, which traces the progenitors of the local subhalos \citep{2015RodriguezG}. It allows one to access the stellar mass assembly over time for each subhalo, as shown in the upper panel of Fig.\,\ref{fig:assembly}. We note that, although a priori it would be possible to recover a DoR from these SFHs as in observations, the reality is that it is not straightforward. Although simulated galaxies present clearly different SFHs, the final assembly time and the fraction of mass assembled within 3\,Gyr (used in INSPIRE to parametrize the DoR), are never as steep as in observations, bringing the DoR values of all simulated CMGs down. This can be seen in Fig.\,\ref{fig:assembly}, comparing the assembly histories of simulated to observed CMGs.

Simulations do not fully produce subhalos as early as seen in observations, taking longer times to complete their build-up, and also presenting very small amounts of star formation until recently. Conversely, observations could be biased towards faster build-up, as the derivation of SFHs strongly depends on the adopted SSP models. This may account for part of the discrepancy with simulations. Another factor that needs to be considered is the stripping process. Environmental effects and tidal forces can affect the estimated mass over time. As a result, the mass obtained in simulations over time is not determined in the same way as in observations, where it is not possible to directly account for the stripping factor. In the end, some CMGs are the result of stripping processes \citep{2025Zhu} and not exclusively the result of the absence of mergers since old times. We highlight these cases in Appendix~\ref{appendixA}, where we compute the SFHs for all the simulated samples alongside their size and mass growth (see Figure \ref{fig:A1}).

To ensure a fairer comparison, we will thus focus on the assembly history during the second phase, since $z=2$, for the simulated sample as described before. For this, we calculate the stellar mass growth fraction, $f_{mass~growth}$, as $(\rm{M}_{\star,z} - \rm{M}_{\star,z=2})/\rm{M}_{\star,z=2}$. This quantifies the amount of stellar mass obtained during the second phase of the evolution of the massive galaxies (from $z=2$ until the $z$ of interest, which is each of the redshift bins in each case).   The upper panel of Fig.\,\ref{fig:assembly} shows the mass assembly over cosmic time for the nine selected compact subhalos at $z=0$, color-coded by this $f_{mass~growth}$.

Hence we use $f_{mass~growth}$ to classify our subhalos into the three classes. For the \textit{relics} class we consider CMGs with less than 20\% of mass growth or accretion, while we are considering the $f_{mass~growth}$ $\leq 10\%$ for the \textit{extreme relics} class. The \textit{non-relic} class are those simulated CMGs with higher mass growth and therefore, accretion fractions (e.g., $\ge 20\%$). The threshold of 10\% for the extreme cases was chosen as it was also applied in \citet{2024Moura} to differentiate relics from other CMGs in TNG50 based on their kinematics. In addition, using three semi-analytical models, \citet{2013Quilis} quantified the fraction and number density of massive galaxies formed at $z>2$ with less than 10\% and 30\% of stellar mass growth. They find that fewer than 2\% of those formed by $z\approx2$ have remained nearly unchanged, accreting no more than 10\% of their mass. The defined extreme threshold is suitable among simulations and also is consistent with the estimated maximum \textit{ex-situ} accretion of NGC\,1277 \citep{2018Beasley}, which is the most extreme case of a relic galaxy in the local Universe today. These threshold values are also directly compatible with the original definition of INSPIRE, although in that case the limit was at 25\%.

\subsection{Stellar mass -- stellar velocity dispersion relations}
\label{sec:kinematics}
One of the main results from INSPIRE is that the DoR strongly correlates with the integrated stellar velocity dispersion. Relic galaxies exhibit higher velocity dispersions compared to non-relics and normal-sized ETGs of similar stellar mass \citep{2024Scognamiglio,2024Maciata,2025Mills}, and thus $\sigma_\star$ can be used to select relic galaxies. This trend is also observed in \citet{2023GrebolTomas}, where Group A galaxies (classified here as relics) display the highest velocity dispersions compared to other groups, standing out as clear outliers from the local scaling relation. This is shown in Fig.~\ref{fig:sigma}, which presents the $\sigma_\star$ and stellar mass relation for both simulated and observed galaxies at each $z$-bin. The dashed black lines denote the median values computed for quiescent subhalos within each redshift, while the black solid line represents the observational stellar mass--velocity dispersion relation from \citet{2016AZahid}. Here, we assume there is not significant evolution with cosmic time for this relation, as \citet{2009Cemarro&Trujillo} showed that spheroidal galaxies with $\rm{M}_\star > 10^{11}\,M_\odot$ at $z \approx 1.6$ exhibit $\sigma_\star$ values similar to those of present-day galaxies. MaNGA and INSPIRE CMGs are color-coded according to their respective groups (see Sec.\,\ref{sec:timescale}), while simulated CMGs are color-coded by their stellar mass growth since $z=2$, as indicated by the colorbar. Simulated CMGs in the $z=2$ bin are left un-colored, as they represent the starting point of the second phase from which the mass growth fraction is computed to assign the classes in simulations. 

Before drawing any conclusions, we must emphasize once more the caveats associated to the intrinsic differences between observations and simulations. \citet{2024Sohnfunction} finds discrepancies in the $\rm{M}_\star-\sigma_\star$ relation, with observed galaxies being on average 40--60\% times higher in velocity dispersion than simulations. They attribute this to the fact that simulated clusters (from TNG300) contain about 60\% fewer subhalos than observed clusters at the same stellar mass limit. Combined with intrinsic differences in the methods to measure this parameter, this results in a simulated velocity dispersion distribution that falls below the observed one with a much smaller scatter, as seen by the dashed line in Fig.~\ref{fig:sigma}. Despite this caveat, in this present work we wish to compare the behavior of the different classes, regardless of their exact $\sigma_\star$ values.

As expected, MaNGA and INSPIRE galaxies show a dispersion in the $\sigma_\star$ at a given stellar mass, with the extreme classes showing much higher $\sigma_\star$ than the rest, and with non-relics following mostly the expected relation. There are no measurements for $\sigma_\star$ available for the VIPERS sample, thus they are not included in this figure. We do not observe a clear trend among the simulated CMGs linking the extreme class to being outliers compared to the non-relic class. Although the highest $\sigma_{\star}$ values at $z=0.7$ are associated with extreme cases, there is considerable scatter. On the contrary, we find that overall our CMGs of all classes tend to exhibit similar and comparable $\sigma_\star$ values. In the future, red nugget data from JWST will allow for a more accurate observational comparison for $z=2$. Naturally, a larger simulated sample is also needed to enable a more robust analysis across all redshifts.

In addition, environmental effects, such as dark matter halo mass and variations in velocity dispersion functions between field and clusters influence velocity dispersion \citep{2018Zahid,2024Sohn}. Intrinsic factors, like black hole mass, also play a role \citep{2000Gebhardt,2015Ferre-mateu}. Therefore, differences in velocity dispersion functions reflect both intrinsic and environmental impacts on galaxy evolution. An additional point is the way that velocity dispersion is calculated in simulations and observations, which differ. The computed does not refer only to the central region of simulated galaxies, which represents a source of incompatibility between the methods.

\begin{figure*}
\centering\includegraphics[width=\textwidth]{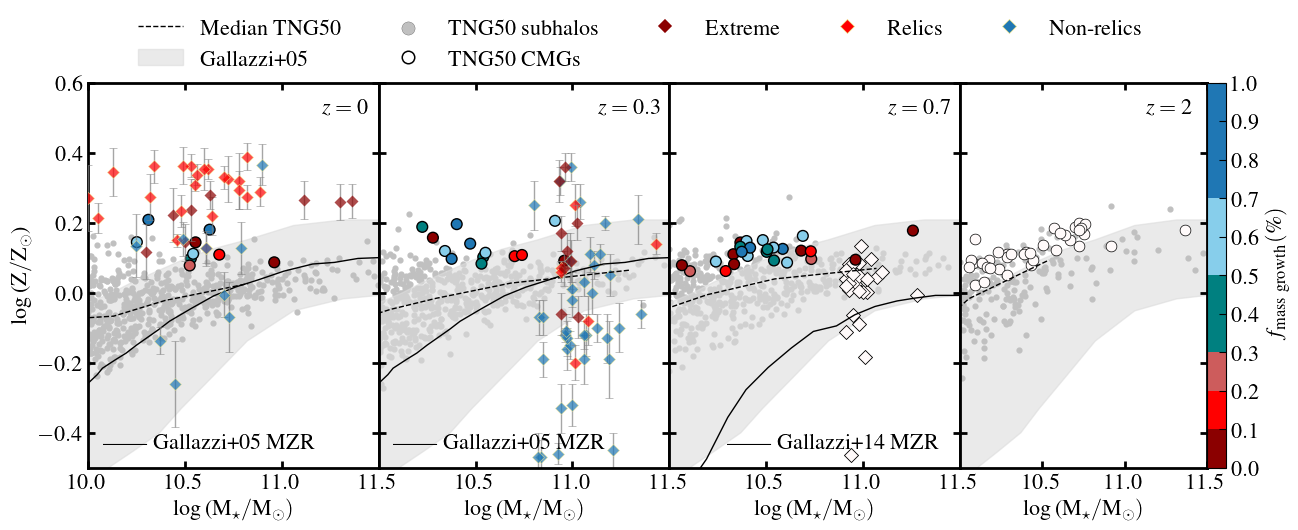}
\caption{Stellar mass -- total stellar metallicity for simulated and observed galaxies. Symbols and colors are as in Fig.\,\ref{fig:sigma}. A shift of $-0.2$\,dex was applied to the stellar metallicities of simulated subhalos following \citet{2019Nelson}. The dashed black line indicates the median stellar metallicity of quiescent subhalos in TNG50. Shaded regions correspond to the local stellar metallicity relations from \citet{2005Gallazzi}; the solid black line follows \citet{2005Gallazzi} for $z=0$ and $z=0.3$, and \citet{2014Gallazzi} for $z=0.7$. Observed relics are more metal-rich than other compact galaxies, deviating from the local mass--metallicity relation, while simulated compact galaxies are consistently more metal-rich than the quiescent population, independent of their SFHs.}
\label{fig:met}
\end{figure*}

\begin{figure}
\centering\includegraphics[width=\columnwidth]{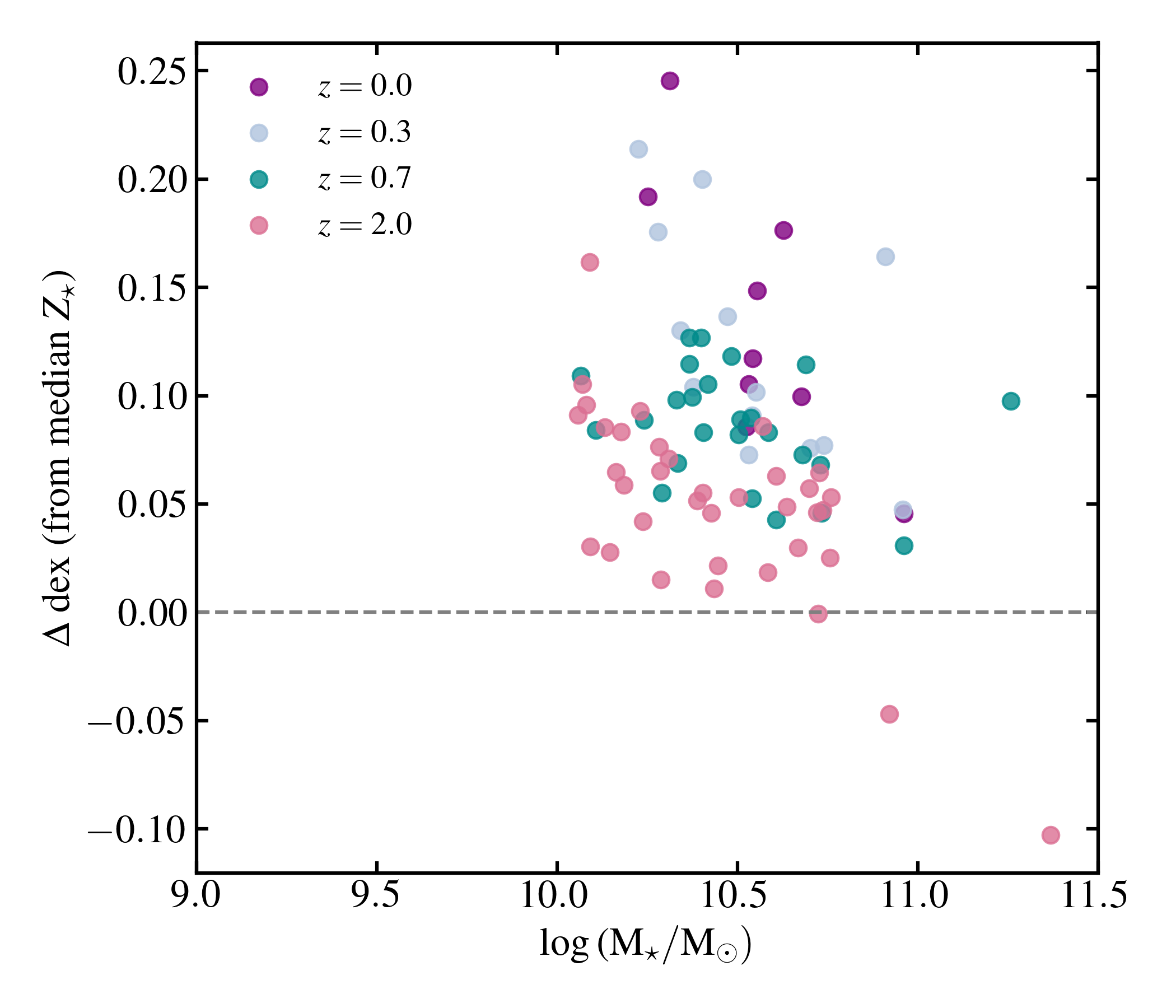}
\caption{Deviation of stellar metallicity from the simulated quiescent median relation as a function of stellar mass for compact galaxies at different redshifts. Each point represents a galaxy, color-coded by redshift. The $\Delta$\,dex values correspond to the offset from the running median of the quiescent population at the same redshift. Gray dashed line marks $\Delta$\,dex = 0, i.e., no deviation from the reference relation. Compact galaxies at all redshifts tend to be more metal-rich than typical quiescent systems at the same mass, with the most pronounced offsets seen at lower redshifts.}
\label{fig:dex_deviation}
\end{figure}
\subsection{Stellar mass -- stellar metallicity relations}
\label{sec:metallicities}

It has been shown that massive relic galaxies exhibit consistently high central metallicities and enhanced [Mg/Fe] abundance ratios, indicative of rapid and early star formation histories \citep{Ferre-Mateu2017, 2023GrebolTomas, 2024Maciata, 2024spiniello, 2025Mills}. When compared to non-relic CMGs, relics stand out for their uniformly elevated chemical signatures and shallower metallicity gradients \citep{Ferre-Mateu2017, 2025Mills}.
These properties are frequently accompanied by evidence of a bottom-heavy IMF, inferred from absorption features sensitive to low-mass stars \citep{2015Martinnavarro, Ferre-Mateu2017, 2023MartinNavarro}. Further analysis reveals a correlation between the DoR, quantifying the fraction of stellar mass formed at $z > 2$, and the IMF slope. Compacts with higher DoR exhibit more bottom-heavy IMFs, suggesting that the conditions prevalent during early cosmic times significantly influence the IMF \citep{2023MartinNavarro, 2024Maciata,2025Mills}. 

Fig.~\ref{fig:met} illustrates the stellar metallicity [M/H], as a function of stellar mass, together with well-established mass\,--\,metallicity relations at different redshifts. We use \citet{2005Gallazzi} for the $z=0$ and $z=0.3$ bins, represented by a black solid line (average value) and the shaded region (intrinsic scatter). For $z=0.7$ we use the mass\,--\,metallicity relations from \citet{2014Gallazzi}, although we still show the local relation as the shaded region. Similarly, we only show the shaded local-region in the $z=2$ panel. The median of the metallicity for all quiescent subhalos in TNG50 (gray circles) at each $z-$bin is shown as a black dashed line. We color-code all the CMGs in simulations and observations following the color scheme of Fig.\,\ref{fig:sigma} for each class: extreme, relics, and non-relics. A shift of $-0.2$\,dex was applied to the stellar metallicities of galaxies with $\rm{M}_\star > 10^{10}\,\rm{M}_\odot$, following \citet{2019Nelson}. Their mock analysis of SDSS-like spectra showed a systematic underestimation of $\sim0.2$ dex in this mass range. Applying this correction ensures a more consistent alignment of metallicity estimates across datasets, although this prevents for a more quantitative comparison in these values.

In terms of the overall trend in Fig.\,\ref{fig:met}, we observe that, at $z = 0$, CMGs classified as extreme and relic in the MaNGA sample all lie above the average metallicity trend of \citet{2005Gallazzi}. A similar behavior is seen in the simulated \textit{extreme} and \textit{relic} galaxies, which are predominantly located above the mean metallicity line for quiescent galaxies in TNG50. In comparison, the selected CMGs occupy a more metal-rich region relative to the general trend of the remaining galaxy population, regardless of their SFHs. This behaviour seems to hold across all redshifts analyzed for simulated CMGs. For galaxies that are not relics or extreme, this clear deviation from the scaling relations might be indicating that in fact are galaxies that have suffered external processes such as stripping, which can remove part of their stellar mass but would keep the central, higher metallicities of the original galaxy (e.g., \citealt{2021Ferre-mateu,2023GrebolTomas}). In fact, our sample of simulated CMGs shows evidence of stellar stripping at different levels, as illustrated by the changing in the mass fraction over time in the upper panel of Fig.~\ref{fig:assembly}. By tracking stellar assembly in the simulations, one can see cases where a decrease in stellar mass is accompanied by a reduction in effective radius, these correlations are detailed in Appendix \ref{fig:A1}. Since our compact galaxy selection is based only on the mass--size relation at a given redshift, this means that an ETG with normal size (non-compact) at early times could undergo strong stripping and end up being classified as a suited CMGs in our selection and analysis. The correlations between galaxy stripping and satellite accretion fractions were investigated in \citet{2024Moura}, which showed that galaxies in high-density environments naturally accrete more satellites and also tend to undergo more stripping.

To investigate the most metal-enhanced systems in each redshift, we next analyze the deviation in stellar metallicity ($\Delta\,\rm{dex}$) of individual galaxies relative to the median mass–metallicity relation at each $z$-bin in Fig.~\ref{fig:met}. The $\Delta\,\rm{dex}$ values were calculated for each galaxy by subtracting the interpolated median metallicity at the corresponding stellar mass. The result can be seen in Fig.\,\ref{fig:dex_deviation}, where we identified extreme outliers as galaxies with $\Delta\,\rm{dex}$ values above the 90th percentile within their respective redshift bins. Although by definition, approximately 10\% of the galaxies are expected to fall in this category, the actual fractions ranged from 11\% to 15\%, depending on the redshift. At low redshift ($z=0$), the threshold for 
extreme deviation was $\Delta$ dex $> 0.203$, with a mean $\Delta$ dex of 0.245 among the extreme objects, indicating the presence of a small but highly enriched population. As redshift increases, the $\Delta$ dex threshold for the 90th percentile gradually decreases (e.g., $\Delta$ dex $> 0.092$ at $z = 2$), and the average deviation among the extreme galaxies becomes more modest ($\sim$0.11 dex). This trend suggests that at earlier cosmic times, even the most metal-rich galaxies deviate less significantly from the population median, possibly reflecting a narrower range of chemical enrichment or more homogeneous evolutionary pathways at high redshift.

\section{Summary and Conclusions}
\label{sec:conclusions}
Massive and compact ETGs show a variety of assembly histories, from very early and fast formation, to late, extended ones. These diverse SFH may be linked to the environment in which the galaxy resides, affecting their dynamics and stellar population properties. In the case of massive relic galaxies, considered as the fossils from the ancient Universe, these CMGs are characterized by uniform stellar ages and elevated stellar metallicities. Given their low-to-inexistent accretion histories, this absence of recent mergers results in a shorter SFH. However, not all relics form as fast but rather show a degree of \textit{relicness} that correlates with the majority of the stellar properties and partially, to the environment. 

In this work, we select CMGs at different cosmic epochs in simulations, applying the same selection criteria used in observations. We focus on redshifts that allow direct comparisons with observational data of CMGs. By applying a uniform classification to analyze both the observational and simulated samples, we ensure a consistent approach that enables a uniform assessment of dynamics and metallicities for samples that share the same initial mass\,--\,size relation. Since we did not aim to identify exact analogues of observed CMGs, but rather to investigate their general properties through their SFHs, this approach allowed us to account for differences in selection, completeness, and modelling, while still enabling global comparisons between simulated and observed populations of CMGs and relic galaxies.

Regarding the mass--size relation, our work yields consistently compact samples at the high-mass end, above $10^{10}\,\mathrm{M_\odot}$, as shown in Fig.~\ref{fig:mas-size}. From this, we conclude that although the stellar mass ranges are not exactly the same at all redshifts (see the Appendix \ref{fig:hist} for mass coverage comparisons), it is reasonable to state that, within the established mass--size relation and selection criteria, our simulated sample is uniform and follows the same trend observed in real CMGs. We also estimated the number density of compact galaxies in TNG50 over cosmic time ($0 < z < 2$), as shown in Fig.~\ref{fig:ndensity}. The values we find are consistently higher than those reported by observations. These results suggest a more abundant compact population in the simulations, but the comparison should be interpreted with caution, since it depends on differences in selection criteria and survey volumes. It is also worth noting that the most extreme relic candidates in the simulation are still less extreme than the prototypical NGC\,1277 (number density $<10^{-6}\,\rm{Mpc^{-3}}$) in their global properties. Furthermore, given their number density in the local Universe, it is unlikely that such rare objects would appear in the simulated volume.

We also explore the $\rm{M}_\star$–$\sigma_\star$ relation for both simulated and observed CMGs across the selected redshifts in Fig.~\ref{fig:sigma}. While observations show that relic galaxies tend to exhibit higher velocity dispersions and stand out as outliers from the local relation, our simulated CMGs present similar $\sigma_\star$ values across different accretion classes, with no strong distinction for the extreme cases. This contrast may be influenced by the limited sample size, intrinsic differences in how $\sigma_\star$ is measured, and environment effects not explored. Nonetheless, the present trends indicate that the simulated CMGs form a dynamically homogeneous population, with comparable $\sigma_\star$ values regardless of their accretion history, or even that $f_{mass~growth}$ does not strongly correlate with the velocity dispersions.

Unlike the case of velocity dispersion, which tends to be underestimated in simulations compared to observations \citep{2024Sohn}, metallicity shows the opposite trend: simulations generally produce higher values when compared to the same observed mass ranges (e.g.,\,\citealt{2018Nelsonn,2017Bahe}). To ensure a fairer comparison, we have applied a shift of $-0.2$\,dex in metallicity for the simulated CMGs (Fig.\,\ref{fig:met}). We find that our CMGs, regardless of their accretion histories, tend to be more metal-rich compared to the median of the quiescent galaxy population in TNG50. This trend is illustrated in Fig.~\ref{fig:dex_deviation}, where local CMGs show a larger deviation from the median metallicity than those at higher redshifts. This suggests the presence of a metallicity gradient, with the deviation in metallicity ($\Delta$\,dex) decreasing at higher redshift.

The first dedicated surveys targeting CMGs, such as INSPIRE and VIPERS, are now providing key observational constraints. Analyses of these systems, both in the local Universe and at higher redshifts, are essential for improving our understanding of their formation and evolution. Looking ahead, future observational efforts will benefit from high-resolution spectroscopic surveys that extend to higher redshifts, as \textit{E}uclid. On the simulation front, initiatives like TNG-Cluster will enable a deeper exploration of how environmental factors influence galaxy dynamics, shape their SFH, and, ultimately, refine the concept of relicness to establish new constraints.


\begin{acknowledgement}
 We thank the anonymous referee for their thoughtful comments, which have helped improve this paper. We thank A. Vazdekis, for his helpful insights and comments during the early stages of this work. MTM acknowledges the Brazilian agencies: Conselho Nacional de Desenvolvimento Científico e Tecnológico (CNPq) through grant 140900/2021-7, and Coordenação de Aperfeiçoamento de Pessoal de Nível Superior (CAPES) through 88887.936624/2024-00. A.F.M. has received support from RYC2021-031099-I and PID2021-123313NA-I00 of MICIN/AEI/10.13039/501100011033/FEDER,UE, NextGenerationEU/PRT. A.C.S. acknowledges support from FAPERGS (grants 23/2551-0001832-2 and 24/2551-0001548-5), CNPq (grants 314301/2021-6, 312940/2025-4, 445231/2024-6, and 404233/2024-4), and CAPES (grant 88887.004427/2024-00). CF ackwoledges funding from CNPq and FAPERGS through grants CNPq-315421/2023-1 and FAPERGS-21/2551-0002025-3.
\end{acknowledgement}

\bibliographystyle{aa}  
\bibliography{ms_relic} 

\begin{thebibliography}{85}
\expandafter\ifx\csname natexlab\endcsname\relax\def\natexlab#1{#1}\fi

\bibitem[{{Alam} {et~al.}(2015){Alam}, {Albareti}, {Allende Prieto}, {Anders}, {Anderson}, {Anderton}, {Andrews}, {Armengaud}, {Aubourg}, {Bailey}, {Basu}, {Bautista}, {Beaton}, {Beers}, {Bender}, {Berlind}, {Beutler}, {Bhardwaj}, {Bird}, {Bizyaev}, {Blake}, {Blanton}, {Blomqvist}, {Bochanski}, {Bolton}, {Bovy}, {Shelden Bradley}, {Brandt}, {Brauer}, {Brinkmann}, {Brown}, {Brownstein}, {Burden}, {Burtin}, {Busca}, {Cai}, {Capozzi}, {Carnero Rosell}, {Carr}, {Carrera}, {Chambers}, {Chaplin}, {Chen}, {Chiappini}, {Chojnowski}, {Chuang}, {Clerc}, {Comparat}, {Covey}, {Croft}, {Cuesta}, {Cunha}, {da Costa}, {Da Rio}, {Davenport}, {Dawson}, {De Lee}, {Delubac}, {Deshpande}, {Dhital}, {Dutra-Ferreira}, {Dwelly}, {Ealet}, {Ebelke}, {Edmondson}, {Eisenstein}, {Ellsworth}, {Elsworth}, {Epstein}, {Eracleous}, {Escoffier}, {Esposito}, {Evans}, {Fan}, {Fern{\'a}ndez-Alvar}, {Feuillet}, {Filiz Ak}, {Finley}, {Finoguenov}, {Flaherty}, {Fleming}, {Font-Ribera}, {Foster}, {Frinchaboy}, {Galbraith-Frew}, {Garc{\'\i}a},
  {Garc{\'\i}a-Hern{\'a}ndez}, {Garc{\'\i}a P{\'e}rez}, {Gaulme}, {Ge}, {G{\'e}nova-Santos}, {Georgakakis}, {Ghezzi}, {Gillespie}, {Girardi}, {Goddard}, {Gontcho}, {Gonz{\'a}lez Hern{\'a}ndez}, {Grebel}, {Green}, {Grieb}, {Grieves}, {Gunn}, {Guo}, {Harding}, {Hasselquist}, {Hawley}, {Hayden}, {Hearty}, {Hekker}, {Ho}, {Hogg}, {Holley-Bockelmann}, {Holtzman}, {Honscheid}, {Huber}, {Huehnerhoff}, {Ivans}, {Jiang}, {Johnson}, {Kinemuchi}, {Kirkby}, {Kitaura}, {Klaene}, {Knapp}, {Kneib}, {Koenig}, {Lam}, {Lan}, {Lang}, {Laurent}, {Le Goff}, {Leauthaud}, {Lee}, {Lee}, {Licquia}, {Liu}, {Long}, {L{\'o}pez-Corredoira}, {Lorenzo-Oliveira}, {Lucatello}, {Lundgren}, {Lupton}, {Mack}, {Mahadevan}, {Maia}, {Majewski}, {Malanushenko}, {Malanushenko}, {Manchado}, {Manera}, {Mao}, {Maraston}, {Marchwinski}, {Margala}, {Martell}, {Martig}, {Masters}, {Mathur}, {McBride}, {McGehee}, {McGreer}, {McMahon}, {M{\'e}nard}, {Menzel}, {Merloni}, {M{\'e}sz{\'a}ros}, {Miller}, {Miralda-Escud{\'e}}, {Miyatake}, {Montero-Dorta}, {More},
  {Morganson}, {Morice-Atkinson}, {Morrison}, {Mosser}, {Muna}, {Myers}, {Nandra}, {Newman}, {Neyrinck}, {Nguyen}, {Nichol}, {Nidever}, {Noterdaeme}, {Nuza}, {O'Connell}, {O'Connell}, {O'Connell}, {Ogando}, {Olmstead}, {Oravetz}, {Oravetz}, {Osumi}, {Owen}, {Padgett}, {Padmanabhan}, {Paegert}, {Palanque-Delabrouille}, \& {Pan}}]{2015Alam}
{Alam}, S., {Albareti}, F.~D., {Allende Prieto}, C., {et~al.} 2015, \apjs, 219, 12

\bibitem[{{Bah{\'e}} {et~al.}(2017){Bah{\'e}}, {Schaye}, {Crain}, {McCarthy}, {Bower}, {Theuns}, {McGee}, \& {Trayford}}]{2017Bahe}
{Bah{\'e}}, Y.~M., {Schaye}, J., {Crain}, R.~A., {et~al.} 2017, \mnras, 464, 508

\bibitem[{{Barro} {et~al.}(2013){Barro}, {Faber}, {P{\'e}rez-Gonz{\'a}lez}, {Koo}, {Williams}, {Kocevski}, {Trump}, {Mozena}, {McGrath}, {van der Wel}, {Wuyts}, {Bell}, {Croton}, {Ceverino}, {Dekel}, {Ashby}, {Cheung}, {Ferguson}, {Fontana}, {Fang}, {Giavalisco}, {Grogin}, {Guo}, {Hathi}, {Hopkins}, {Huang}, {Koekemoer}, {Kartaltepe}, {Lee}, {Newman}, {Porter}, {Primack}, {Ryan}, {Rosario}, {Somerville}, {Salvato}, \& {Hsu}}]{2013Barro}
{Barro}, G., {Faber}, S.~M., {P{\'e}rez-Gonz{\'a}lez}, P.~G., {et~al.} 2013, \apj, 765, 104

\bibitem[{{Beasley} {et~al.}(2018){Beasley}, {Trujillo}, {Leaman}, \& {Montes}}]{2018Beasley}
{Beasley}, M.~A., {Trujillo}, I., {Leaman}, R., \& {Montes}, M. 2018, \nat, 555, 483

\bibitem[{{Bundy} {et~al.}(2015){Bundy}, {Bershady}, {Law}, {Yan}, {Drory}, {MacDonald}, {Wake}, {Cherinka}, {S{\'a}nchez-Gallego}, {Weijmans}, {Thomas}, {Tremonti}, {Masters}, {Coccato}, {Diamond-Stanic}, {Arag{\'o}n-Salamanca}, {Avila-Reese}, {Badenes}, {Falc{\'o}n-Barroso}, {Belfiore}, {Bizyaev}, {Blanc}, {Bland-Hawthorn}, {Blanton}, {Brownstein}, {Byler}, {Cappellari}, {Conroy}, {Dutton}, {Emsellem}, {Etherington}, {Frinchaboy}, {Fu}, {Gunn}, {Harding}, {Johnston}, {Kauffmann}, {Kinemuchi}, {Klaene}, {Knapen}, {Leauthaud}, {Li}, {Lin}, {Maiolino}, {Malanushenko}, {Malanushenko}, {Mao}, {Maraston}, {McDermid}, {Merrifield}, {Nichol}, {Oravetz}, {Pan}, {Parejko}, {Sanchez}, {Schlegel}, {Simmons}, {Steele}, {Steinmetz}, {Thanjavur}, {Thompson}, {Tinker}, {van den Bosch}, {Westfall}, {Wilkinson}, {Wright}, {Xiao}, \& {Zhang}}]{2015ABundy}
{Bundy}, K., {Bershady}, M.~A., {Law}, D.~R., {et~al.} 2015, \apj, 798, 7

\bibitem[{{Carnall} {et~al.}(2024){Carnall}, {Cullen}, {McLure}, {McLeod}, {Begley}, {Donnan}, {Dunlop}, {Shapley}, {Rowlands}, {Almaini}, {Arellano-C{\'o}rdova}, {Barrufet}, {Cimatti}, {Ellis}, {Grogin}, {Hamadouche}, {Illingworth}, {Koekemoer}, {Leung}, {Lovell}, {P{\'e}rez-Gonz{\'a}lez}, {Santini}, {Stanton}, \& {Wild}}]{2024Carnall}
{Carnall}, A.~C., {Cullen}, F., {McLure}, R.~J., {et~al.} 2024, \mnras, 534, 325

\bibitem[{{Carnall} {et~al.}(2023){Carnall}, {McLeod}, {McLure}, {Dunlop}, {Begley}, {Cullen}, {Donnan}, {Hamadouche}, {Jewell}, {Jones}, {Pollock}, \& {Wild}}]{2023Carnall}
{Carnall}, A.~C., {McLeod}, D.~J., {McLure}, R.~J., {et~al.} 2023, \mnras, 520, 3974

\bibitem[{{Cebri{\'a}n} \& {Trujillo}(2014)}]{2014mariacebrian}
{Cebri{\'a}n}, M. \& {Trujillo}, I. 2014, \mnras, 444, 682

\bibitem[{{Cenarro} \& {Trujillo}(2009)}]{2009Cemarro&Trujillo}
{Cenarro}, A.~J. \& {Trujillo}, I. 2009, \apjl, 696, L43

\bibitem[{{Chabrier}(2003)}]{2003Chabrier}
{Chabrier}, G. 2003, \pasp, 115, 763

\bibitem[{{Charbonnier} {et~al.}(2017){Charbonnier}, {Huertas-Company}, {Gon{\c{c}}alves}, {Men{\'e}ndez-Delmestre}, {Bundy}, {Galliano}, {Moraes}, {Makler}, {Pereira}, {Erben}, {Hildebrandt}, {Shan}, {Caminha}, {Grossi}, \& {Riguccini}}]{2017Charbonnier}
{Charbonnier}, A., {Huertas-Company}, M., {Gon{\c{c}}alves}, T.~S., {et~al.} 2017, \mnras, 469, 4523

\bibitem[{{Clerici} {et~al.}(2024){Clerici}, {Schnorr-M{\"u}ller}, {Trevisan}, \& {Ricci}}]{2024Clerici}
{Clerici}, K.~S., {Schnorr-M{\"u}ller}, A., {Trevisan}, M., \& {Ricci}, T.~V. 2024, \mnras, 531, 1034

\bibitem[{{Crain} {et~al.}(2015){Crain}, {Schaye}, {Bower}, {Furlong}, {Schaller}, {Theuns}, {Dalla Vecchia}, {Frenk}, {McCarthy}, {Helly}, {Jenkins}, {Rosas-Guevara}, {White}, \& {Trayford}}]{2015Crain}
{Crain}, R.~A., {Schaye}, J., {Bower}, R.~G., {et~al.} 2015, \mnras, 450, 1937

\bibitem[{{D'Ago} {et~al.}(2023){D'Ago}, {Spiniello}, {Coccato}, {Tortora}, {La Barbera}, {Arnaboldi}, {Bevacqua}, {Ferr{\'e}-Mateu}, {Gallazzi}, {Hartke}, {Hunt}, {Mart{\'\i}n-Navarro}, {Napolitano}, {Pulsoni}, {Radovich}, {Saracco}, {Scognamiglio}, \& {Zibetti}}]{2023ADago}
{D'Ago}, G., {Spiniello}, C., {Coccato}, L., {et~al.} 2023, \aap, 672, A17

\bibitem[{{Damjanov} {et~al.}(2015){Damjanov}, {Geller}, {Zahid}, \& {Hwang}}]{2015Damjanov}
{Damjanov}, I., {Geller}, M.~J., {Zahid}, H.~J., \& {Hwang}, H.~S. 2015, \apj, 806, 158

\bibitem[{{Damjanov} {et~al.}(2014){Damjanov}, {Hwang}, {Geller}, \& {Chilingarian}}]{2014Damjanov}
{Damjanov}, I., {Hwang}, H.~S., {Geller}, M.~J., \& {Chilingarian}, I. 2014, \apj, 793, 39

\bibitem[{{Damjanov} {et~al.}(2009){Damjanov}, {McCarthy}, {Abraham}, {Glazebrook}, {Yan}, {Mentuch}, {Le Borgne}, {Savaglio}, {Crampton}, {Murowinski}, {Juneau}, {Carlberg}, {J{\o}rgensen}, {Roth}, {Chen}, \& {Marzke}}]{2009Damjanov}
{Damjanov}, I., {McCarthy}, P.~J., {Abraham}, R.~G., {et~al.} 2009, \apj, 695, 101

\bibitem[{{Dolag} {et~al.}(2009){Dolag}, {Borgani}, {Murante}, \& {Springel}}]{2009Dolag}
{Dolag}, K., {Borgani}, S., {Murante}, G., \& {Springel}, V. 2009, \mnras, 399, 497

\bibitem[{{Edge} {et~al.}(2013){Edge}, {Sutherland}, {Kuijken}, {Driver}, {McMahon}, {Eales}, \& {Emerson}}]{2013Edge}
{Edge}, A., {Sutherland}, W., {Kuijken}, K., {et~al.} 2013, The Messenger, 154, 32

\bibitem[{{Ferr{\'e}-Mateu} {et~al.}(2021){Ferr{\'e}-Mateu}, {Durr{\'e}}, {Forbes}, {Romanowsky}, {Alabi}, {Brodie}, \& {McDermid}}]{2021Ferre-mateu}
{Ferr{\'e}-Mateu}, A., {Durr{\'e}}, M., {Forbes}, D.~A., {et~al.} 2021, \mnras, 503, 5455

\bibitem[{{Ferr{\'e}-Mateu} {et~al.}(2015){Ferr{\'e}-Mateu}, {Mezcua}, {Trujillo}, {Balcells}, \& {van den Bosch}}]{2015Ferre-mateu}
{Ferr{\'e}-Mateu}, A., {Mezcua}, M., {Trujillo}, I., {Balcells}, M., \& {van den Bosch}, R. C.~E. 2015, \apj, 808, 79

\bibitem[{{Ferr{\'e}-Mateu} {et~al.}(2012){Ferr{\'e}-Mateu}, {Vazdekis}, {Trujillo}, {S{\'a}nchez-Bl{\'a}zquez}, {Ricciardelli}, \& {de la Rosa}}]{2012Ferre-mateu}
{Ferr{\'e}-Mateu}, A., {Vazdekis}, A., {Trujillo}, I., {et~al.} 2012, \mnras, 423, 632

\bibitem[{Ferré-Mateu {et~al.}(2017)Ferré-Mateu, Forbes, Romanowsky, Janz, \& Dixon}]{Ferre-Mateu2017}
Ferré-Mateu, A., Forbes, D.~A., Romanowsky, A.~J., Janz, J., \& Dixon, C. 2017, Monthly Notices of the Royal Astronomical Society, 473, 1819

\bibitem[{Flores-Freitas {et~al.}(2022)Flores-Freitas, Chies-Santos, Furlanetto, De Rossi, Ferreira, Zenocratti, \& Alamo-Martínez}]{Floresfreitas2022}
Flores-Freitas, R., Chies-Santos, A.~L., Furlanetto, C., {et~al.} 2022, Monthly Notices of the Royal Astronomical Society, 512, 245

\bibitem[{{Gallazzi} {et~al.}(2014){Gallazzi}, {Bell}, {Zibetti}, {Brinchmann}, \& {Kelson}}]{2014Gallazzi}
{Gallazzi}, A., {Bell}, E.~F., {Zibetti}, S., {Brinchmann}, J., \& {Kelson}, D.~D. 2014, \apj, 788, 72

\bibitem[{{Gallazzi} {et~al.}(2005){Gallazzi}, {Charlot}, {Brinchmann}, {White}, \& {Tremonti}}]{2005Gallazzi}
{Gallazzi}, A., {Charlot}, S., {Brinchmann}, J., {White}, S. D.~M., \& {Tremonti}, C.~A. 2005, \mnras, 362, 41

\bibitem[{{Garcia} {et~al.}(2024){Garcia}, {Torrey}, {Grasha}, {Hernquist}, {Ellison}, {Zovaro}, {Hemler}, {Nelson}, \& {Kewley}}]{2024Garcia}
{Garcia}, A.~M., {Torrey}, P., {Grasha}, K., {et~al.} 2024, \mnras, 529, 3342

\bibitem[{{Gebhardt} {et~al.}(2000){Gebhardt}, {Bender}, {Bower}, {Dressler}, {Faber}, {Filippenko}, {Green}, {Grillmair}, {Ho}, {Kormendy}, {Lauer}, {Magorrian}, {Pinkney}, {Richstone}, \& {Tremaine}}]{2000Gebhardt}
{Gebhardt}, K., {Bender}, R., {Bower}, G., {et~al.} 2000, \apjl, 539, L13

\bibitem[{{Genel} {et~al.}(2018){Genel}, {Nelson}, {Pillepich}, {Springel}, {Pakmor}, {Weinberger}, {Hernquist}, {Naiman}, {Vogelsberger}, {Marinacci}, \& {Torrey}}]{2018Genel}
{Genel}, S., {Nelson}, D., {Pillepich}, A., {et~al.} 2018, \mnras, 474, 3976

\bibitem[{{Gr{\`e}bol-Tom{\`a}s} {et~al.}(2023){Gr{\`e}bol-Tom{\`a}s}, {Ferr{\'e}-Mateu}, \& {Dom{\'\i}nguez-S{\'a}nchez}}]{2023GrebolTomas}
{Gr{\`e}bol-Tom{\`a}s}, P., {Ferr{\'e}-Mateu}, A., \& {Dom{\'\i}nguez-S{\'a}nchez}, H. 2023, \mnras, 526, 4024

\bibitem[{{Guidi} {et~al.}(2016){Guidi}, {Scannapieco}, {Walcher}, \& {Gallazzi}}]{2016Guidi}
{Guidi}, G., {Scannapieco}, C., {Walcher}, J., \& {Gallazzi}, A. 2016, \mnras, 462, 2046

\bibitem[{{Ito} {et~al.}(2024){Ito}, {Valentino}, {Brammer}, {Faisst}, {Gillman}, {G{\'o}mez-Guijarro}, {Gould}, {Heintz}, {Ilbert}, {Jespersen}, {Kokorev}, {Kubo}, {Magdis}, {McPartland}, {Onodera}, {Rizzo}, {Tanaka}, {Toft}, {Vijayan}, {Weaver}, {Whitaker}, \& {Wright}}]{2024valentino}
{Ito}, K., {Valentino}, F., {Brammer}, G., {et~al.} 2024, \apj, 964, 192

\bibitem[{{Lisiecki} {et~al.}(2023){Lisiecki}, {Ma{\l}ek}, {Siudek}, {Pollo}, {Krywult}, {Karska}, \& {Junais}}]{2023Lisiecki}
{Lisiecki}, K., {Ma{\l}ek}, K., {Siudek}, M., {et~al.} 2023, \aap, 669, A95

\bibitem[{{Long} {et~al.}(2024){Long}, {Antwi-Danso}, {Lambrides}, {Lovell}, {de la Vega}, {Valentino}, {Zavala}, {Casey}, {Wilkins}, {Yung}, {Arrabal Haro}, {Bagley}, {Bisigello}, {Chworowsky}, {Cooper}, {Cooper}, {Cooray}, {Croton}, {Dickinson}, {Finkelstein}, {Franco}, {Gould}, {Hirschmann}, {Hutchison}, {Kartaltepe}, {Kocevski}, {Koekemoer}, {Lucas}, {McKinney}, {Nere}, {Papovich}, {P{\'e}rez-Gonz{\'a}lez}, {Pirzkal}, \& {Santini}}]{2024Long}
{Long}, A.~S., {Antwi-Danso}, J., {Lambrides}, E.~L., {et~al.} 2024, \apj, 970, 68

\bibitem[{{Maksymowicz-Maciata} {et~al.}(2024){Maksymowicz-Maciata}, {Spiniello}, {Mart{\'\i}n-Navarro}, {Ferr{\'e}-Mateu}, {Bevacqua}, {Cappellari}, {D'Ago}, {Tortora}, {Arnaboldi}, {Hartke}, {Napolitano}, {Saracco}, \& {Scognamiglio}}]{2024Maciata}
{Maksymowicz-Maciata}, M., {Spiniello}, C., {Mart{\'\i}n-Navarro}, I., {et~al.} 2024, \mnras, 531, 2864

\bibitem[{{Marinacci} {et~al.}(2018){Marinacci}, {Vogelsberger}, {Pakmor}, {Torrey}, {Springel}, {Hernquist}, {Nelson}, {Weinberger}, {Pillepich}, {Naiman}, \& {Genel}}]{2018Marinacci}
{Marinacci}, F., {Vogelsberger}, M., {Pakmor}, R., {et~al.} 2018, \mnras, 480, 5113

\bibitem[{{Mart{\'\i}n-Navarro} {et~al.}(2015){Mart{\'\i}n-Navarro}, {La Barbera}, {Vazdekis}, {Ferr{\'e}-Mateu}, {Trujillo}, \& {Beasley}}]{2015Martinnavarro}
{Mart{\'\i}n-Navarro}, I., {La Barbera}, F., {Vazdekis}, A., {et~al.} 2015, \mnras, 451, 1081

\bibitem[{{Mart{\'\i}n-Navarro} {et~al.}(2023){Mart{\'\i}n-Navarro}, {Spiniello}, {Tortora}, {Coccato}, {D'Ago}, {Ferr{\'e}-Mateu}, {Pulsoni}, {Hartke}, {Arnaboldi}, {Hunt}, {Napolitano}, {Scognamiglio}, \& {Spavone}}]{2023MartinNavarro}
{Mart{\'\i}n-Navarro}, I., {Spiniello}, C., {Tortora}, C., {et~al.} 2023, \mnras, 521, 1408

\bibitem[{{Mills} {et~al.}(2025){Mills}, {Spiniello}, {Sergeyev}, {Tortora}, {Khramtsov}, {D'Ago}, {Maksymowicz-Maciata}, {Benedetti}, {Ferr{\'e}-Mateu}, {Cappellari}, {Davies}, {Hartke}, \& {Rosen}}]{2025Mills}
{Mills}, J., {Spiniello}, C., {Sergeyev}, A., {et~al.} 2025, arXiv e-prints, arXiv:2501.16126

\bibitem[{{Moura} {et~al.}(2024){Moura}, {Chies-Santos}, {Furlanetto}, {Zhu}, \& {Canossa-Gosteinski}}]{2024Moura}
{Moura}, M.~T., {Chies-Santos}, A.~L., {Furlanetto}, C., {Zhu}, L., \& {Canossa-Gosteinski}, M.~A. 2024, \mnras, 528, 353

\bibitem[{{Naab} {et~al.}(2009){Naab}, {Johansson}, \& {Ostriker}}]{2009Naab}
{Naab}, T., {Johansson}, P.~H., \& {Ostriker}, J.~P. 2009, \apjl, 699, L178

\bibitem[{{Naiman} {et~al.}(2018){Naiman}, {Pillepich}, {Springel}, {Ramirez-Ruiz}, {Torrey}, {Vogelsberger}, {Pakmor}, {Nelson}, {Marinacci}, {Hernquist}, {Weinberger}, \& {Genel}}]{2018Naiman}
{Naiman}, J.~P., {Pillepich}, A., {Springel}, V., {et~al.} 2018, \mnras, 477, 1206

\bibitem[{{Nanayakkara} {et~al.}(2025){Nanayakkara}, {Glazebrook}, {Schreiber}, {Chittenden}, {Brammer}, {Esdaile}, {Jacobs}, {Kacprzak}, {Kawinwanichakij}, {Kimmig}, {Labbe}, {Lagos}, {Marchesini}, {Mart{\`\i}nez-Mar{\`\i}n}, {Marsan}, {Oesch}, {Papovich}, {Remus}, \& {Tran}}]{2025Nana}
{Nanayakkara}, T., {Glazebrook}, K., {Schreiber}, C., {et~al.} 2025, \apj, 981, 78

\bibitem[{{Nelson} {et~al.}(2018{\natexlab{a}}){Nelson}, {Pillepich}, {Springel}, {Weinberger}, {Hernquist}, {Pakmor}, {Genel}, {Torrey}, {Vogelsberger}, {Kauffmann}, {Marinacci}, \& {Naiman}}]{2018Nelson}
{Nelson}, D., {Pillepich}, A., {Springel}, V., {et~al.} 2018{\natexlab{a}}, \mnras, 475, 624

\bibitem[{{Nelson} {et~al.}(2018{\natexlab{b}}){Nelson}, {Pillepich}, {Springel}, {Weinberger}, {Hernquist}, {Pakmor}, {Genel}, {Torrey}, {Vogelsberger}, {Kauffmann}, {Marinacci}, \& {Naiman}}]{2018Nelsonn}
{Nelson}, D., {Pillepich}, A., {Springel}, V., {et~al.} 2018{\natexlab{b}}, \mnras, 475, 624

\bibitem[{{Nelson} {et~al.}(2019){Nelson}, {Springel}, {Pillepich}, {Rodriguez-Gomez}, {Torrey}, {Genel}, {Vogelsberger}, {Pakmor}, {Marinacci}, {Weinberger}, {Kelley}, {Lovell}, {Diemer}, \& {Hernquist}}]{2019Nelson}
{Nelson}, D., {Springel}, V., {Pillepich}, A., {et~al.} 2019, Computational Astrophysics and Cosmology, 6, 2

\bibitem[{{Oser} {et~al.}(2010){Oser}, {Ostriker}, {Naab}, {Johansson}, \& {Burkert}}]{2010OserOstriker}
{Oser}, L., {Ostriker}, J.~P., {Naab}, T., {Johansson}, P.~H., \& {Burkert}, A. 2010, \apj, 725, 2312

\bibitem[{{Pakmor} {et~al.}(2011){Pakmor}, {Bauer}, \& {Springel}}]{2011Pakmor}
{Pakmor}, R., {Bauer}, A., \& {Springel}, V. 2011, \mnras, 418, 1392

\bibitem[{{Panter} {et~al.}(2008){Panter}, {Jimenez}, {Heavens}, \& {Charlot}}]{2008Panter}
{Panter}, B., {Jimenez}, R., {Heavens}, A.~F., \& {Charlot}, S. 2008, \mnras, 391, 1117

\bibitem[{{Peralta de Arriba} {et~al.}(2016){Peralta de Arriba}, {Quilis}, {Trujillo}, {Cebri{\'a}n}, \& {Balcells}}]{2016PeraltaArriba}
{Peralta de Arriba}, L., {Quilis}, V., {Trujillo}, I., {Cebri{\'a}n}, M., \& {Balcells}, M. 2016, \mnras, 461, 156

\bibitem[{{Pillepich} {et~al.}(2019){Pillepich}, {Nelson}, {Springel}, {Pakmor}, {Torrey}, {Weinberger}, {Vogelsberger}, {Marinacci}, {Genel}, {van der Wel}, \& {Hernquist}}]{2019Pillepich}
{Pillepich}, A., {Nelson}, D., {Springel}, V., {et~al.} 2019, \mnras, 490, 3196

\bibitem[{{Pillepich} {et~al.}(2018){Pillepich}, {Springel}, {Nelson}, {Genel}, {Naiman}, {Pakmor}, {Hernquist}, {Torrey}, {Vogelsberger}, {Weinberger}, \& {Marinacci}}]{2018Pillepich}
{Pillepich}, A., {Springel}, V., {Nelson}, D., {et~al.} 2018, \mnras, 473, 4077

\bibitem[{{Planck Collaboration} {et~al.}(2016){Planck Collaboration}, {Ade, P. A. R.}, {Aghanim, N.}, {Arnaud, M.}, {Ashdown, M.}, {Aumont, J.}, {Baccigalupi, C.}, {Banday, A. J.}, {Barreiro, R. B.}, {Bartlett, J. G.}, {Bartolo, N.}, {Battaner, E.}, {Battye, R.}, {Benabed, K.}, {Beno\^{\i}t, A.}, {Benoit-L\'evy, A.}, {Bernard, J.-P.}, {Bersanelli, M.}, {Bielewicz, P.}, {Bock, J. J.}, {Bonaldi, A.}, {Bonavera, L.}, {Bond, J. R.}, {Borrill, J.}, {Bouchet, F. R.}, {Boulanger, F.}, {Bucher, M.}, {Burigana, C.}, {Butler, R. C.}, {Calabrese, E.}, {Cardoso, J.-F.}, {Catalano, A.}, {Challinor, A.}, {Chamballu, A.}, {Chary, R.-R.}, {Chiang, H. C.}, {Chluba, J.}, {Christensen, P. R.}, {Church, S.}, {Clements, D. L.}, {Colombi, S.}, {Colombo, L. P. L.}, {Combet, C.}, {Coulais, A.}, {Crill, B. P.}, {Curto, A.}, {Cuttaia, F.}, {Danese, L.}, {Davies, R. D.}, {Davis, R. J.}, {de Bernardis, P.}, {de Rosa, A.}, {de Zotti, G.}, {Delabrouille, J.}, {D\'esert, F.-X.}, {Di Valentino, E.}, {Dickinson, C.}, {Diego, J. M.}, {Dolag,
  K.}, {Dole, H.}, {Donzelli, S.}, {Dor\'e, O.}, {Douspis, M.}, {Ducout, A.}, {Dunkley, J.}, {Dupac, X.}, {Efstathiou, G.}, {Elsner, F.}, {En\ss{}lin, T. A.}, {Eriksen, H. K.}, {Farhang, M.}, {Fergusson, J.}, {Finelli, F.}, {Forni, O.}, {Frailis, M.}, {Fraisse, A. A.}, {Franceschi, E.}, {Frejsel, A.}, {Galeotta, S.}, {Galli, S.}, {Ganga, K.}, {Gauthier, C.}, {Gerbino, M.}, {Ghosh, T.}, {Giard, M.}, {Giraud-H\'eraud, Y.}, {Giusarma, E.}, {Gjerl\o{}w, E.}, {Gonz\'alez-Nuevo, J.}, {G\'orski, K. M.}, {Gratton, S.}, {Gregorio, A.}, {Gruppuso, A.}, {Gudmundsson, J. E.}, {Hamann, J.}, {Hansen, F. K.}, {Hanson, D.}, {Harrison, D. L.}, {Helou, G.}, {Henrot-Versill\'e, S.}, {Hern\'andez-Monteagudo, C.}, {Herranz, D.}, {Hildebrandt, S. R.}, {Hivon, E.}, {Hobson, M.}, {Holmes, W. A.}, {Hornstrup, A.}, {Hovest, W.}, {Huang, Z.}, {Huffenberger, K. M.}, {Hurier, G.}, {Jaffe, A. H.}, {Jaffe, T. R.}, {Jones, W. C.}, {Juvela, M.}, {Keih\"anen, E.}, {Keskitalo, R.}, {Kisner, T. S.}, {Kneissl, R.}, {Knoche, J.}, {Knox, L.},
  {Kunz, M.}, {Kurki-Suonio, H.}, {Lagache, G.}, {L\"ahteenm\"aki, A.}, {Lamarre, J.-M.}, {Lasenby, A.}, {Lattanzi, M.}, {Lawrence, C. R.}, {Leahy, J. P.}, {Leonardi, R.}, {Lesgourgues, J.}, {Levrier, F.}, {Lewis, A.}, {Liguori, M.}, {Lilje, P. B.}, {Linden-V\o{}rnle, M.}, {L\'opez-Caniego, M.}, {Lubin, P. M.}, {Mac\'{\i}as-P\'erez, J. F.}, {Maggio, G.}, {Maino, D.}, {Mandolesi, N.}, {Mangilli, A.}, {Marchini, A.}, {Maris, M.}, {Martin, P. G.}, {Martinelli, M.}, {Mart\'{\i}nez-Gonz\'alez, E.}, {Masi, S.}, {Matarrese, S.}, {McGehee, P.}, {Meinhold, P. R.}, {Melchiorri, A.}, {Melin, J.-B.}, {Mendes, L.}, {Mennella, A.}, {Migliaccio, M.}, {Millea, M.}, {Mitra, S.}, {Miville-Desch\^enes, M.-A.}, {Moneti, A.}, {Montier, L.}, {Morgante, G.}, {Mortlock, D.}, {Moss, A.}, {Munshi, D.}, {Murphy, J. A.}, {Naselsky, P.}, {Nati, F.}, {Natoli, P.}, {Netterfield, C. B.}, {N\o{}rgaard-Nielsen, H. U.}, {Noviello, F.}, {Novikov, D.}, {Novikov, I.}, {Oxborrow, C. A.}, {Paci, F.}, {Pagano, L.}, {Pajot, F.}, {Paladini, R.},
  {Paoletti, D.}, {Partridge, B.}, {Pasian, F.}, {Patanchon, G.}, {Pearson, T. J.}, {Perdereau, O.}, {Perotto, L.}, {Perrotta, F.}, {Pettorino, V.}, {Piacentini, F.}, {Piat, M.}, {Pierpaoli, E.}, {Pietrobon, D.}, {Plaszczynski, S.}, {Pointecouteau, E.}, {Polenta, G.}, {Popa, L.}, {Pratt, G. W.}, {Pr\'ezeau, G.}, {Prunet, S.}, {Puget, J.-L.}, {Rachen, J. P.}, {Reach, W. T.}, {Rebolo, R.}, {Reinecke, M.}, {Remazeilles, M.}, {Renault, C.}, {Renzi, A.}, {Ristorcelli, I.}, {Rocha, G.}, {Rosset, C.}, {Rossetti, M.}, {Roudier, G.}, {Rouill\'e d\'{}Orfeuil, B.}, {Rowan-Robinson, M.}, {Rubi\~no-Mart\'{\i}n, J. A.}, {Rusholme, B.}, {Said, N.}, {Salvatelli, V.}, {Salvati, L.}, {Sandri, M.}, {Santos, D.}, {Savelainen, M.}, {Savini, G.}, {Scott, D.}, {Seiffert, M. D.}, {Serra, P.}, {Shellard, E. P. S.}, {Spencer, L. D.}, {Spinelli, M.}, {Stolyarov, V.}, {Stompor, R.}, {Sudiwala, R.}, {Sunyaev, R.}, {Sutton, D.}, {Suur-Uski, A.-S.}, {Sygnet, J.-F.}, {Tauber, J. A.}, {Terenzi, L.}, {Toffolatti, L.}, {Tomasi, M.}, {Tristram,
  M.}, {Trombetti, T.}, {Tucci, M.}, {Tuovinen, J.}, {T\"urler, M.}, {Umana, G.}, {Valenziano, L.}, {Valiviita, J.}, {Van Tent, F.}, {Vielva, P.}, {Villa, F.}, {Wade, L. A.}, {Wandelt, B. D.}, {Wehus, I. K.}, {White, M.}, {White, S. D. M.}, {Wilkinson, A.}, {Yvon, D.}, {Zacchei, A.}, \& {Zonca, A.}}]{plank2016}
{Planck Collaboration}, {Ade, P. A. R.}, {Aghanim, N.}, {et~al.} 2016, A\&A, 594, A13

\bibitem[{{Poggianti} {et~al.}(2013){Poggianti}, {Calvi}, {Bindoni}, {D'Onofrio}, {Moretti}, {Valentinuzzi}, {Fasano}, {Fritz}, {De Lucia}, {Vulcani}, {Bettoni}, {Gullieuszik}, \& {Omizzolo}}]{2013Poggianti}
{Poggianti}, B.~M., {Calvi}, R., {Bindoni}, D., {et~al.} 2013, \apj, 762, 77

\bibitem[{{Quilis} \& {Trujillo}(2013)}]{2013Quilis}
{Quilis}, V. \& {Trujillo}, I. 2013, \apjl, 773, L8

\bibitem[{{Rodriguez-Gomez} {et~al.}(2015){Rodriguez-Gomez}, {Genel}, {Vogelsberger}, {Sijacki}, {Pillepich}, {Sales}, {Torrey}, {Snyder}, {Nelson}, {Springel}, {Ma}, \& {Hernquist}}]{2015RodriguezG}
{Rodriguez-Gomez}, V., {Genel}, S., {Vogelsberger}, M., {et~al.} 2015, \mnras, 449, 49

\bibitem[{{Roy} {et~al.}(2018){Roy}, {Napolitano}, {La Barbera}, {Tortora}, {Getman}, {Radovich}, {Capaccioli}, {Brescia}, {Cavuoti}, {Longo}, {Raj}, {Puddu}, {Covone}, {Amaro}, {Vellucci}, {Grado}, {Kuijken}, {Verdoes Kleijn}, \& {Valentijn}}]{2018Roy}
{Roy}, N., {Napolitano}, N.~R., {La Barbera}, F., {et~al.} 2018, \mnras, 480, 1057

\bibitem[{{Schaye} {et~al.}(2015){Schaye}, {Crain}, {Bower}, {Furlong}, {Schaller}, {Theuns}, {Dalla Vecchia}, {Frenk}, {McCarthy}, {Helly}, {Jenkins}, {Rosas-Guevara}, {White}, {Baes}, {Booth}, {Camps}, {Navarro}, {Qu}, {Rahmati}, {Sawala}, {Thomas}, \& {Trayford}}]{2015Schaye}
{Schaye}, J., {Crain}, R.~A., {Bower}, R.~G., {et~al.} 2015, \mnras, 446, 521

\bibitem[{{Scodeggio} {et~al.}(2018{\natexlab{a}}){Scodeggio}, {Guzzo}, {Garilli}, {Granett}, {Bolzonella}, {de la Torre}, {Abbas}, {Adami}, {Arnouts}, {Bottini}, {Cappi}, {Coupon}, {Cucciati}, {Davidzon}, {Franzetti}, {Fritz}, {Iovino}, {Krywult}, {Le Brun}, {Le F{\`e}vre}, {Maccagni}, {Ma{\l}ek}, {Marchetti}, {Marulli}, {Polletta}, {Pollo}, {Tasca}, {Tojeiro}, {Vergani}, {Zanichelli}, {Bel}, {Branchini}, {De Lucia}, {Ilbert}, {McCracken}, {Moutard}, {Peacock}, {Zamorani}, {Burden}, {Fumana}, {Jullo}, {Marinoni}, {Mellier}, {Moscardini}, \& {Percival}}]{2018Scodeggio}
{Scodeggio}, M., {Guzzo}, L., {Garilli}, B., {et~al.} 2018{\natexlab{a}}, \aap, 609, A84

\bibitem[{{Scodeggio} {et~al.}(2018{\natexlab{b}}){Scodeggio}, {Guzzo}, {Garilli}, {Granett}, {Bolzonella}, {de la Torre}, {Abbas}, {Adami}, {Arnouts}, {Bottini}, {Cappi}, {Coupon}, {Cucciati}, {Davidzon}, {Franzetti}, {Fritz}, {Iovino}, {Krywult}, {Le Brun}, {Le F{\`e}vre}, {Maccagni}, {Ma{\l}ek}, {Marchetti}, {Marulli}, {Polletta}, {Pollo}, {Tasca}, {Tojeiro}, {Vergani}, {Zanichelli}, {Bel}, {Branchini}, {De Lucia}, {Ilbert}, {McCracken}, {Moutard}, {Peacock}, {Zamorani}, {Burden}, {Fumana}, {Jullo}, {Marinoni}, {Mellier}, {Moscardini}, \& {Percival}}]{2018vimos}
{Scodeggio}, M., {Guzzo}, L., {Garilli}, B., {et~al.} 2018{\natexlab{b}}, \aap, 609, A84

\bibitem[{{Scognamiglio} {et~al.}(2024){Scognamiglio}, {Spiniello}, {Radovich}, {Tortora}, {Napolitano}, {Li}, {Maturi}, {Maksymowicz-Maciata}, {Cappellari}, {Arnaboldi}, {Bevacqua}, {Coccato}, {D'Ago}, {Feng}, {Ferr{\'e}-Mateu}, {Hartke}, {Mart{\'\i}n-Navarro}, \& {Pulsoni}}]{2024Scognamiglio}
{Scognamiglio}, D., {Spiniello}, C., {Radovich}, M., {et~al.} 2024, \mnras, 534, 1597

\bibitem[{{Scognamiglio} {et~al.}(2020){Scognamiglio}, {Tortora}, {Spavone}, {Spiniello}, {Napolitano}, {D'Ago}, {La Barbera}, {Getman}, {Roy}, {Raj}, {Radovich}, {Brescia}, {Cavuoti}, {Koopmans}, {Kuijken}, {Longo}, \& {Petrillo}}]{2020AScognamiglio}
{Scognamiglio}, D., {Tortora}, C., {Spavone}, M., {et~al.} 2020, \apj, 893, 4

\bibitem[{{Siudek} {et~al.}(2023){Siudek}, {Lisiecki}, {Krywult}, {Donevski}, {Haines}, {Karska}, {Ma{\l}ek}, {Moutard}, \& {Pollo}}]{2023Siudek}
{Siudek}, M., {Lisiecki}, K., {Krywult}, J., {et~al.} 2023, \mnras, 523, 4294

\bibitem[{{Sohn} {et~al.}(2024{\natexlab{a}}){Sohn}, {Geller}, {Borrow}, \& {Vogelsberger}}]{2024Sohnfunction}
{Sohn}, J., {Geller}, M.~J., {Borrow}, J., \& {Vogelsberger}, M. 2024{\natexlab{a}}, \apj, 974, 26

\bibitem[{{Sohn} {et~al.}(2024{\natexlab{b}}){Sohn}, {Geller}, {Borrow}, \& {Vogelsberger}}]{2024Sohn}
{Sohn}, J., {Geller}, M.~J., {Borrow}, J., \& {Vogelsberger}, M. 2024{\natexlab{b}}, \apj, 964, 178

\bibitem[{{Spiniello} {et~al.}(2024){Spiniello}, {D'Ago}, {Coccato}, {Hartke}, {Tortora}, {Ferr{\'e}-Mateu}, {Pulsoni}, {Cappellari}, {Maksymowicz-Maciata}, {Arnaboldi}, {Bevacqua}, {Gallazzi}, {Hunt}, {La Barbera}, {Mart{\'\i}n-Navarro}, {Napolitano}, {Radovich}, {Saracco}, {Scognamiglio}, {Spavone}, \& {Zibetti}}]{2024spiniello}
{Spiniello}, C., {D'Ago}, G., {Coccato}, L., {et~al.} 2024, \mnras, 527, 8793

\bibitem[{{Spiniello} {et~al.}(2021{\natexlab{a}}){Spiniello}, {Tortora}, {D'Ago}, {Coccato}, {La Barbera}, {Ferr{\'e}-Mateu}, {Napolitano}, {Spavone}, {Scognamiglio}, {Arnaboldi}, {Gallazzi}, {Hunt}, {Moehler}, {Radovich}, \& {Zibetti}}]{2021INSPIREproject}
{Spiniello}, C., {Tortora}, C., {D'Ago}, G., {et~al.} 2021{\natexlab{a}}, \aap, 646, A28

\bibitem[{{Spiniello} {et~al.}(2021{\natexlab{b}}){Spiniello}, {Tortora}, {D'Ago}, {Coccato}, {La Barbera}, {Ferr{\'e}-Mateu}, {Napolitano}, {Spavone}, {Scognamiglio}, {Arnaboldi}, {Gallazzi}, {Hunt}, {Moehler}, {Radovich}, \& {Zibetti}}]{2021Spiniello}
{Spiniello}, C., {Tortora}, C., {D'Ago}, G., {et~al.} 2021{\natexlab{b}}, \aap, 646, A28

\bibitem[{{Springel}(2010)}]{2010Springel}
{Springel}, V. 2010, \mnras, 401, 791

\bibitem[{{Springel} \& {Hernquist}(2003)}]{2003Sringel}
{Springel}, V. \& {Hernquist}, L. 2003, \mnras, 339, 289

\bibitem[{{Springel} {et~al.}(2018){Springel}, {Pakmor}, {Pillepich}, {Weinberger}, {Nelson}, {Hernquist}, {Vogelsberger}, {Genel}, {Torrey}, {Marinacci}, \& {Naiman}}]{2018Springel}
{Springel}, V., {Pakmor}, R., {Pillepich}, A., {et~al.} 2018, \mnras, 475, 676

\bibitem[{{Stringer} {et~al.}(2015){Stringer}, {Trujillo}, {Dalla Vecchia}, \& {Martinez-Valpuesta}}]{2015Stringer}
{Stringer}, M., {Trujillo}, I., {Dalla Vecchia}, C., \& {Martinez-Valpuesta}, I. 2015, \mnras, 449, 2396

\bibitem[{{Tortora} {et~al.}(2018){Tortora}, {Napolitano}, {Spavone}, {La Barbera}, {D'Ago}, {Spiniello}, {Kuijken}, {Roy}, {Raj}, {Cavuoti}, {Brescia}, {Longo}, {Pota}, {Petrillo}, {Radovich}, {Getman}, {Koopmans}, {Trujillo}, {Verdoes Kleijn}, {Capaccioli}, {Grado}, {Covone}, {Scognamiglio}, {Blake}, {Glazebrook}, {Joudaki}, {Lidman}, \& {Wolf}}]{2018Tortora}
{Tortora}, C., {Napolitano}, N.~R., {Spavone}, M., {et~al.} 2018, \mnras, 481, 4728

\bibitem[{{Trujillo} {et~al.}(2009){Trujillo}, {Cenarro}, {de Lorenzo-C{\'a}ceres}, {Vazdekis}, {de la Rosa}, \& {Cava}}]{2009Trujillo}
{Trujillo}, I., {Cenarro}, A.~J., {de Lorenzo-C{\'a}ceres}, A., {et~al.} 2009, \apjl, 692, L118

\bibitem[{{Trujillo} {et~al.}(2014){Trujillo}, {Ferr{\'e}-Mateu}, {Balcells}, {Vazdekis}, \& {S{\'a}nchez-Bl{\'a}zquez}}]{2014Trujillo}
{Trujillo}, I., {Ferr{\'e}-Mateu}, A., {Balcells}, M., {Vazdekis}, A., \& {S{\'a}nchez-Bl{\'a}zquez}, P. 2014, \apjl, 780, L20

\bibitem[{van~der Wel {et~al.}(2014)van~der Wel, Franx, van Dokkum, Skelton, Momcheva, Whitaker, Brammer, Bell, Rix, Wuyts, Ferguson, Holden, Barro, Koekemoer, Chang, McGrath, Häussler, Dekel, Behroozi, Fumagalli, Leja, Lundgren, Maseda, Nelson, Wake, Patel, Labbé, Faber, Grogin, \& Kocevski}]{vanderWel2014}
van~der Wel, A., Franx, M., van Dokkum, P.~G., {et~al.} 2014, The Astrophysical Journal, 788, 28

\bibitem[{{van Dokkum} {et~al.}(2015){van Dokkum}, {Nelson}, {Franx}, {Oesch}, {Momcheva}, {Brammer}, {F{\"o}rster Schreiber}, {Skelton}, {Whitaker}, {van der Wel}, {Bezanson}, {Fumagalli}, {Illingworth}, {Kriek}, {Leja}, \& {Wuyts}}]{2015ApJVanDokkun}
{van Dokkum}, P.~G., {Nelson}, E.~J., {Franx}, M., {et~al.} 2015, \apj, 813, 23

\bibitem[{{Wellons} {et~al.}(2015){Wellons}, {Torrey}, {Ma}, {Rodriguez-Gomez}, {Vogelsberger}, {Kriek}, {van Dokkum}, {Nelson}, {Genel}, {Pillepich}, {Springel}, {Sijacki}, {Snyder}, {Nelson}, {Sales}, \& {Hernquist}}]{2015Wellons}
{Wellons}, S., {Torrey}, P., {Ma}, C.-P., {et~al.} 2015, \mnras, 449, 361

\bibitem[{Yıldırım {et~al.}(2017)Yıldırım, van~den Bosch, van~de Ven, Martín-Navarro, Walsh, Husemann, Gültekin, \& Gebhardt}]{Yildrim2017}
Yıldırım, A., van~den Bosch, R. C.~E., van~de Ven, G., {et~al.} 2017, Monthly Notices of the Royal Astronomical Society, 468, 4216

\bibitem[{{Zahid} {et~al.}(2016){Zahid}, {Geller}, {Fabricant}, \& {Hwang}}]{2016AZahid}
{Zahid}, H.~J., {Geller}, M.~J., {Fabricant}, D.~G., \& {Hwang}, H.~S. 2016, \apj, 832, 203

\bibitem[{{Zahid} {et~al.}(2018){Zahid}, {Sohn}, \& {Geller}}]{2018Zahid}
{Zahid}, H.~J., {Sohn}, J., \& {Geller}, M.~J. 2018, \apj, 859, 96

\bibitem[{{Zhu} {et~al.}(2025){Zhu}, {Chies-Santos}, {Moura}, \& {Shi}}]{2025Zhu}
{Zhu}, L., {Chies-Santos}, A.~L., {Moura}, M.~T., \& {Shi}, H. 2025, \aap, 698, A195

\bibitem[{{Zhu} {et~al.}(2022){Zhu}, {Pillepich}, {van de Ven}, {Leaman}, {Hernquist}, {Nelson}, {Pakmor}, {Vogelsberger}, \& {Zhang}}]{2022ZhuLing}
{Zhu}, L., {Pillepich}, A., {van de Ven}, G., {et~al.} 2022, \aap, 660, A20

\bibitem[{{Zibetti} {et~al.}(2020){Zibetti}, {Gallazzi}, {Hirschmann}, {Consolandi}, {Falc{\'o}n-Barroso}, {van de Ven}, \& {Lyubenova}}]{2020Zibetti}
{Zibetti}, S., {Gallazzi}, A.~R., {Hirschmann}, M., {et~al.} 2020, \mnras, 491, 3562

\bibitem[{{Zolotov} {et~al.}(2015){Zolotov}, {Dekel}, {Mandelker}, {Tweed}, {Inoue}, {DeGraf}, {Ceverino}, {Primack}, {Barro}, \& {Faber}}]{2015Zolotov}
{Zolotov}, A., {Dekel}, A., {Mandelker}, N., {et~al.} 2015, \mnras, 450, 2327

\end{thebibliography}
\appendix
\begin{appendix} 
\onecolumn
\section{Additional Figures}
\label{appendixA}

\begin{figure*}
\centering
\includegraphics[scale=0.5]{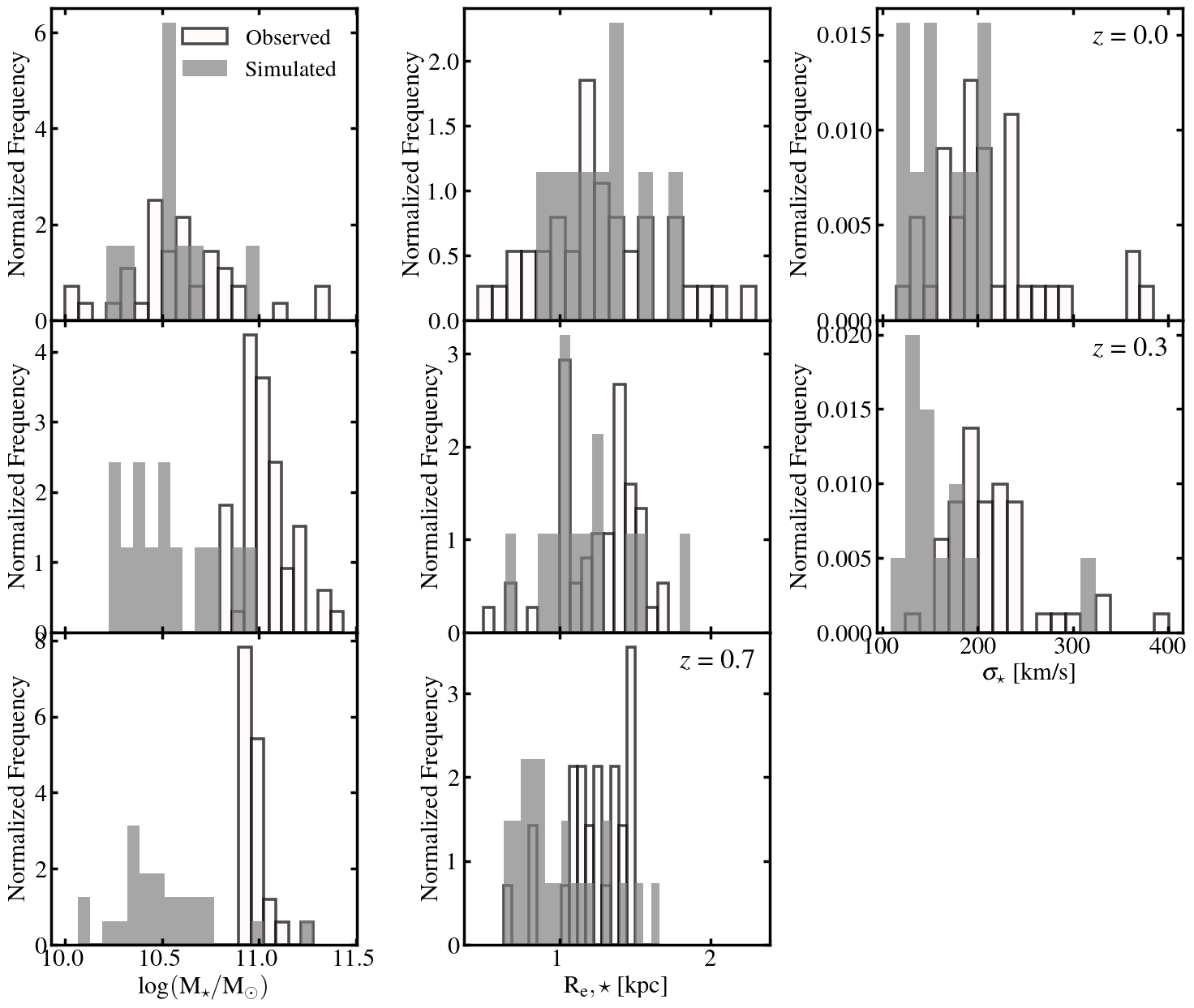}
\caption{Normalized distributions of stellar mass as $\log\rm{(M_\star/M}_\odot)$, stellar velocity dispersion ($\sigma_\star$), and effective radius ($R_{\rm e,\star}$) for observed and simulated galaxies at the three distinctive redshift bins in this study: $z = 0.0$, $z = 0.3$, and $z = 0.7$. Observed data are shown in open histograms and simulated data in grey filled ones. The observational stellar velocity dispersion at $z = 0.7$ is omitted due to the lack of available measurements in the VIPERS dataset.}
\label{fig:hist}
\end{figure*}

\begin{figure*}
\centering
\includegraphics[width=\linewidth]{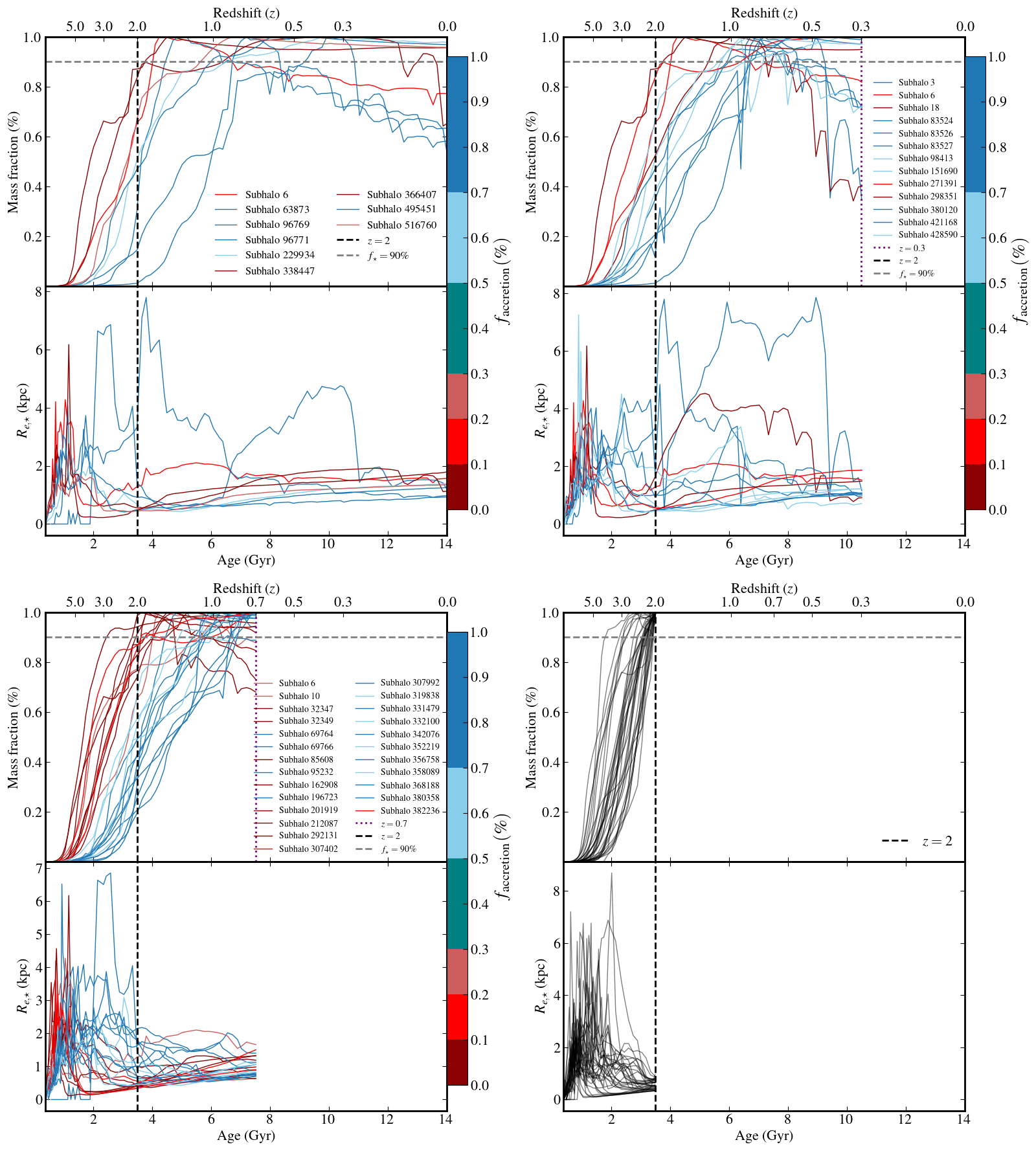}
\caption{Stellar mass and size assembly of simulated compact galaxies at four redshifts: $z=0$ (upper left), $z=0.3$ (upper right), $z=0.7$ (lower left), and $z=2$ (lower right). Each panel displays the evolution of stellar mass fraction and effective radius over time. All panels are color-coded by the stellar mass accreted since $z=2$, except for the $z=2$ panel, which remains uncolored. The vertical dashed black line marks $z=2$, while the dashed purple line indicates the redshift corresponding to each sample. The horizontal grey line denotes the 90\% stellar mass fraction threshold at the given redshift.}.
\label{fig:A1}
\end{figure*}


\begin{table*}
\caption{Sample selection across redshift bins ($z=0.0,0.3,0.7\,\rm{and}\,2.0$) and their associated properties. Columns correspond to galaxy ID, stellar mass, effective radius, velocity dispersion, and stellar metallicity, respectively. Each redshift section is separated by the horizontal lines.}
\label{tab:prop}
\centering
\scriptsize

\begin{tabular}{lccccc}
\hline\hline
ID & $\log\,(\rm{M_{\star}/M_\odot)}$ & $R_{\mathrm{e}, \star}~\mathrm{(kpc)}$ & $\sigma_\star~\mathrm{(km/s)}$ & $\log\,(Z/Z_\odot)$ \\

\hline 
6      & 10.67742  & 1.57192 & 173.36389 & 0.46022 \\
63873  & 10.62765  & 1.18470 & 205.14405 & 0.53157 \\
96769  & 10.31156  & 0.97799 & 153.24827 & 0.55978 \\
96771  & 10.25199  & 0.92846 & 150.37505 & 0.49603 \\
229934 & 10.53091  & 1.38919 & 131.70610 & 0.44949 \\
338447 & 10.55485  & 1.13194 & 200.01434 & 0.49540 \\
366407 & 10.96106  & 1.79427 & 192.13632 & 0.43832 \\
495451 & 10.54248  & 1.25734 & 124.25655 & 0.46265 \\
516760 & 10.52482  & 1.35546 & 128.08775 & 0.42921 \\

\hline 

3      & 10.91010 & 1.15609 & 308.99641 & 0.55684 \\
6      & 10.70199 & 1.52225 & 178.34175 & 0.45583 \\
18     & 10.27999 & 0.91345 & 146.18132 & 0.50869 \\
83524  & 10.47237 & 0.97377 & 177.15799 & 0.49162 \\
83526  & 10.40290 & 1.05759 & 149.11661 & 0.54734 \\
83527  & 10.34209 & 1.04081 & 137.80829 & 0.47065 \\
98413  & 10.22512 & 0.70943 & 142.53642 & 0.53954 \\
151690 & 10.53988 & 1.22348 & 136.75015 & 0.45353 \\
271391 & 10.73954 & 1.86750 & 159.68944 & 0.45928 \\
298351 & 10.95795 & 1.48336 & 194.04102 & 0.44285 \\
380120 & 10.37668 & 1.02889 & 108.14479 & 0.44831 \\
421168 & 10.55122 & 1.08043 & 131.22266 & 0.46560 \\
428590 & 10.53073 & 1.22236 & 135.39551 & 0.43444 \\

\hline 
3      & 10.91010 & 1.15609 & 308.99641  & 0.55684 \\
6      & 10.73280 & 1.66287 & 174.96576  & 0.44772 \\
10     & 10.10680 & 0.63980 & 118.98294  & 0.41300 \\
32347  & 10.68057 & 1.19071 & 200.57047  & 0.47166 \\
32349  & 10.36704 & 0.88089 & 169.07697  & 0.49501 \\
69764  & 10.58559 & 1.11954 & 170.04167  & 0.47674 \\
69766  & 10.39863 & 0.69827 & 144.20665  & 0.49913 \\
85608  & 10.33075 & 0.75832 & 173.54291  & 0.46146 \\
95232  & 10.40461 & 0.80310 & 155.21078  & 0.45634 \\
162908 & 11.25744 & 1.33067 & 249.84113  & 0.52902 \\
196723 & 10.68927 & 1.40411 & 158.08151  & 0.51379 \\
201919 & 10.06522 & 0.63362 & 149.39571  & 0.43007 \\
212087 & 10.96119 & 1.32342 & 197.83391  & 0.44574 \\
292131 & 10.33447 & 0.98671 & 139.53224  & 0.43292 \\
307402 & 10.73041 & 1.50590 & 158.97818  & 0.46981 \\
307992 & 10.48288 & 0.72208 & 141.87112  & 0.50173 \\
319838 & 10.36645 & 0.82238 & 127.00477  & 0.48281 \\
331479 & 10.60800 & 1.06284 & 139.04102  & 0.43753 \\
332100 & 10.50272 & 1.16515 & 131.25653  & 0.46815 \\
342076 & 10.53783 & 0.80008 & 144.77278  & 0.48038 \\
352219 & 10.50714 & 0.87671 & 133.38252  & 0.47553 \\
356758 & 10.41717 & 0.89029 & 125.82673  & 0.48009 \\
358089 & 10.54044 & 1.05756 & 141.44965  & 0.44350 \\
368188 & 10.24044 & 0.74469 & 114.71716  & 0.44056 \\
380358 & 10.37398 & 0.76760 & 118.27165  & 0.46852 \\
382236 & 10.29095 & 0.90583 & 120.09289  & 0.41338 \\

\hline 
8072   & 10.17779 & 0.81109 & 132.76302 & 0.44578 \\
8074   & 10.08940 & 0.52072 & 142.62384 & 0.50421 \\
21554  & 11.36775 & 0.95274 & 290.41492 & 0.52920 \\
25822  & 10.75562 & 0.45346 & 250.35625 & 0.51844 \\
25823  & 10.50258 & 0.35823 & 186.0383  & 0.48915 \\
37160  & 10.08147 & 0.65918 & 135.05478 & 0.43645 \\
39746  & 10.72408 & 0.50987 & 272.70322 & 0.48543 \\
44316  & 10.28330 & 0.38085 & 145.4977  & 0.46281 \\
59076  & 10.57042 & 0.64005 & 219.22235 & 0.53735 \\
60751  & 10.28589 & 0.52278 & 152.2319  & 0.45214 \\
60752  & 10.40343 & 0.86432 & 197.14742 & 0.46863 \\
66743  & 10.92041 & 0.49061 & 225.53642 & 0.48374 \\
71129  & 10.58467 & 0.75011 & 171.89828 & 0.47311 \\
72910  & 10.69890 & 0.42178 & 220878.0  & 0.53782 \\
75669  & 10.07040 & 0.40367 & 134.6908  & 0.44356 \\
81249  & 10.72613 & 0.35504 & 208.33542 & 0.55124 \\
84667  & 10.75894 & 0.32787 & 209.02861 & 0.54724 \\
86525  & 10.66774 & 0.63950 & 198.35374 & 0.50318 \\
89720  & 10.73668 & 0.34106 & 194.14516 & 0.53612 \\
93214  & 10.72038 & 0.77223 & 186.27931 & 0.53155 \\
102999 & 10.43463 & 0.54126 & 178.73729 & 0.43161 \\
108487 & 10.44632 & 0.89635 & 160.59843 & 0.44488 \\
109728 & 10.63620 & 0.70984 & 169.98878 & 0.51493 \\
114745 & 10.38873 & 0.51332 & 148.78058 & 0.46179 \\
115247 & 10.60713 & 0.67607 & 178.68933 & 0.52271 \\
122298 & 10.09123 & 0.64571 & 157.73535 & 0.37330 \\
126923 & 10.23794 & 0.56154 & 130.08905 & 0.41818 \\
129239 & 10.13235 & 0.54785 & 131.99518 & 0.43763 \\
133326 & 10.42714 & 0.67791 & 152.81583 & 0.46482 \\
137192 & 10.28751 & 0.80278 & 145.81677 & 0.40253 \\
137657 & 10.14620 & 0.72992 & 140.83842 & 0.38316 \\
155422 & 10.30855 & 0.44564 & 146.67227 & 0.46297 \\
176382 & 10.16368 & 0.77723 & 128.47844 & 0.42409 \\
178254 & 10.18421 & 0.80828 & 128.35857 & 0.42282 \\
183362 & 10.22955 & 0.33900 & 138.37595 & 0.46720 \\
238669 & 10.05729 & 0.34887 & 121.40987 & 0.42616 \\
\hline
\end{tabular}
\label{tab:ids}
\end{table*}

\end{appendix}
\end{document}